\documentclass{sig-alternate} 
\usepackage{mathptmx} 

\newcommand{\ignore}[1]{}
\usepackage{fancyhdr}
\usepackage[normalem]{ulem}
\usepackage[hyphens]{url}
\usepackage{microtype}
\usepackage{enumitem}

\usepackage{comment}
\usepackage{listings}

\setlist[itemize,1]{itemsep=0.5pt,partopsep=0pt,parsep=\parskip, topsep=2pt, leftmargin=10pt,}
\setlist[enumerate,1]{itemsep=0.5pt,partopsep=0pt,parsep=\parskip, topsep=2pt, leftmargin=10pt,}

\usepackage{color}
\definecolor{bg}{rgb}{0.9,0.9,0.9}
\definecolor{dkgreen}{rgb}{0,0.6,0}
\definecolor{gray}{rgb}{0.5,0.5,0.5}
\definecolor{mauve}{rgb}{0.58,0,0.82}

\usepackage{subcaption}

\usepackage[bookmarks=true,breaklinks=true,letterpaper=true,colorlinks,linkcolor=black,citecolor=blue,urlcolor=black]{hyperref}

\newcommand{\iscasubmissionnumber}{NaN}

\fancypagestyle{firstpage}{
  \fancyhf{}

  \fancyhead[C]{\normalsize{ISCA 2018 Submission
      \textbf{\#\iscasubmissionnumber} \\ Confidential Draft: DO NOT DISTRIBUTE}} 
  \fancyfoot[C]{\thepage}
}  

\title{\vspace{-1em}Best-Effort FPGA Programming: \\A Few Steps Can Go a Long Way\vspace{-2.5em}} 
\author{
 Jason Cong \\ \email{cong@cs.ucla.edu} \and
 Zhenman Fang \\ \email{zhenman@cs.ucla.edu} \and
 Yuchen Hao \\ \email{haoyc@cs.ucla.edu} \and
 Peng Wei \\ \email{peng.wei.prc@cs.ucla.edu} \and
 Cody Hao Yu \\ \email{hyu@cs.ucla.edu} \and
 Chen Zhang \\ \email{chenzhang2015@ucla.edu} \and
 Peipei Zhou \\ \email{memoryzpp@cs.ucla.edu}
 \\
 {Center for Domain-Specific Computing, University of California, Los Angeles}
}

\begin{document}
\maketitle
\pagestyle{plain}

\begin{abstract}
FPGA-based heterogeneous architectures are attracting ever-increasing attention from both academia and industry in an attempt to advance computational capabilities and energy efficiency in today's datacenters.
These architectures provide programmers with the ability to customize their hardware accelerators for flexible acceleration of many workloads.
Nonetheless, such advantages come at the cost of sacrificing programmability.
FPGA vendors and researchers attempt to improve the programmability through high-level synthesis (HLS) technologies that can directly generate hardware circuits from high-level language descriptions.
However, reading through recent publications on FPGA designs using HLS, one often gets the impression that FPGA programming is still \emph{hard} in that it leaves programmers to explore a very large design space with many possible combinations of HLS optimization strategies, and it often requires intimate knowledge of the FPGA architecture to select the right combination to achieve the best result.

In this paper we make two important observations and contributions. First, we demonstrate a rather surprising result: FPGA programming can be made \emph{easy} by following a simple best-effort guideline of five refinement (sub)steps using HLS. 
We show that for a broad class of accelerator benchmarks from MachSuite, the proposed best-effort guideline improves the FPGA accelerator performance by 42-29,030x. Compared to the baseline CPU performance, the FPGA accelerator performance is improved from an average 292.5x slowdown to an average 34.4x speedup. 
Moreover, we show that the refinement steps in the best-effort guideline, consisting of explicit data caching, customized pipelining, processing element duplication, computation/communication overlapping and scratchpad reorganization, correspond well to the best practice guidelines for multicore CPU programming. 
We plan to open-source our best-effort programming guideline and design templates to the community.
Although our best-effort guideline may not always lead to the optimal solution, it substantially simplifies the FPGA programming effort, and will greatly support the wide adoption of FPGA-based acceleration by the software programming community. 

\end{abstract}

\section{Introduction} \label{sec:intro}

Due to power and energy constraints, conventional general-purpose processors are no longer able to sustain the performance and energy improvement in commercial datacenters.
To overcome the inefficiencies of homogeneous multicore systems, heterogeneous architectures that feature specialized hardware accelerators have been widely considered as a promising paradigm.
Field programmable gate arrays (FPGAs), which offer the potential of orders-of-magnitude performance/watt gains for a broad class of applications while retaining reconfigurability, attract increasing attention as a mainstream acceleration technology.
For example, both Microsoft and Baidu have incorporated FPGA-based accelerators in their datacenters to accelerate large-scale production workloads such as search engines~\cite{msft-catapult-isca14,msft-micro16} and neural networks~\cite{baidu-sda-hotchips14,msft-cnn-hotchips15}.
Moreover, with the \$16.7 billion acquisition of Altera, Intel recently announced the Heterogeneous Architecture Research Platform (HARP)~\cite{harp},
aiming to provide an FPGA and a Xeon processor in a single semiconductor package. Predictions have been made that as much as 30\% of datacenter servers will have FPGAs by 2020~\cite{harp-30}. 
This suggests that FPGAs could become a common component in future servers and could play an important role as primary computing resources~\cite{eriko2016fpl}.

A major challenge in riding on the free performance lunch of FPGA is programmability.
FPGA programming is generally recognized as an RTL (register-transfer level) design practice, which requires notable hardware expertise in designing accelerator microarchitectures such as controls, data paths, and finite state machines~\cite{rmm}. 
This makes it prohibitive to most software programmers.
It is even more challenging when the mainstream algorithm in an application domain is constantly evolving; i.e., an algorithm may have already been obsolete during the development process of its hardware accelerator.

\begin{figure}[!t]
  \centering
  \includegraphics[width=\columnwidth]{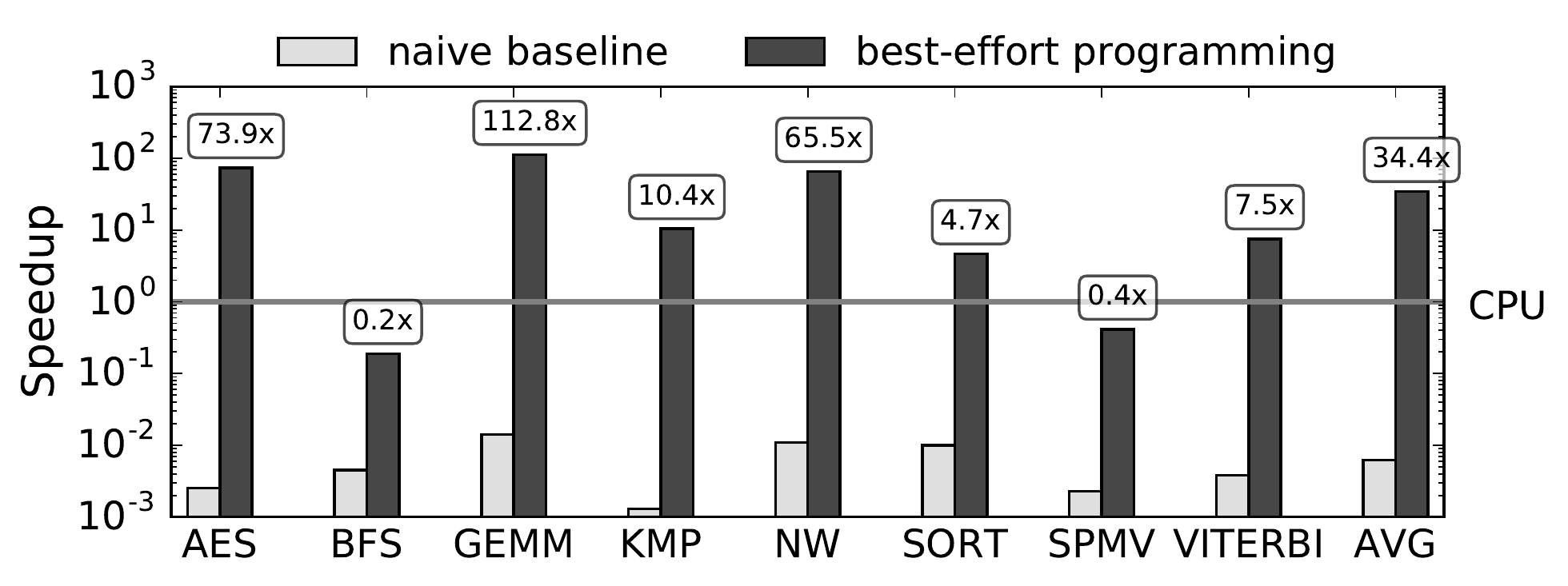}
  \caption{How far can best-effort FPGA programming go: from worse than 200x slowdown to more than 30x speedup over single-thread CPU implementations.}
  \label{fig:speedup-overall}
\end{figure}

Decades of research has been focusing on improving FPGA programmability. High-Level Synthesis (HLS)~\cite{jason2011HLS} can derive high-quality accelerator designs directly from high-level behavioral descriptions, saving programmers from extensive hand-coding in RTL and manual tuning. 
State-of-the-art HLS tools such as the Xilinx Vivado HLS~\cite{vivado} and Intel FPGA SDK for OpenCL~\cite{aocl} allow computational kernels to be described in C and OpenCL, which can then be compiled and synthesized into FPGA accelerators. Using these tools, programmers can easily write synthesizable code or convert existing software implementations into FPGA accelerators. 

However, a simple push-button process is far from producing high-performance FPGA accelerators.
To demonstrate this, we compare the performance of a single-thread CPU with FPGA accelerators that are directly generated from the same software code~\footnote{We only made necessary interface changes in order to synthesize them. Figure~\ref{fig:baseline} shows the AES example.} of benchmarks from MachSuite~\cite{machsuite}.
As shown in Figure~\ref{fig:speedup-overall}, these naive FPGA accelerators are more than 200x slower than the original software implementations running on a Xeon CPU core (see Section \ref{sec:expsetup} for detailed experimental setup), which defeats our purpose of accelerating these computational kernels in the first place.
%
To improve the quality of HLS-generated accelerators, many prior studies~\cite{Cong-dac12, Pham-date15, Pouchet-fpga13, Hagiescu-dac09, Cong-fpga14, liu-fccm15, liu-fccm16, Venkat-pldi15, raghu-asplos, cgpa, elasticflow, weisz2013fpga, Wang-dac13, Su-fpga16, Cilardo-taco15} focus on proposing enhancements to HLS languages to express certain hardware structures.
Nonetheless, most studies require the understanding of hardware le in order for programmers to direct HLS tools to generate the right hardware structure. Also, choosing the right combination of optimization strategies among an exponential set of candidates is non-trivial even for experienced accelerator designers~\cite{fpga-opencl}. As a result, programmers often get the impression that designing high-performance FPGA accelerators is still much harder compared to software programming.

In this paper we aim to address the following questions: \textbf{(1) Can mainstream software programmers make high-quality FPGA accelerators with affordable programming efforts? (2) What are the best-effort guidelines for them to achieve this goal?}

To answer these questions, we attempt to apply various HLS optimizations to the ported computational kernels from MachSuite~\cite{machsuite}. 
Encouragingly, we observe that the following best-effort programming with a small set of programmer-accessible HLS optimizations can produce quite compelling results: as shown in Figure~\ref{fig:speedup-overall}, the best-effort programming improves the accelerator performance by an average of 6702x over the naive baseline, while outperforming a Xeon CPU core by 34.4x on average. 
In our best-effort guideline, we use data-driven refinement to iteratively optimize the accelerator design: in each refinement iteration, we pinpoint the performance bottleneck and apply a small set of HLS optimizations listed in Table~\ref{tab:strategy}.
These HLS optimizations include explicit data caching through batch processing and data tiling, customized pipelining, processing element (PE) duplication, double buffering and scratchpad reorganization.
To provide more insights, we also summarize the CPU programming counterpart, example code pointer, and performance impact of each optimization technique in Table~\ref{tab:strategy}.

\begin{table}[t]
\centering
\caption{Summary of optimization strategies.}
\label{tab:strategy}
{
\scriptsize
\begin{tabular}{|l|l|l|l|}
\hline
HLS Optimization          & \begin{tabular}[c]{@{}l@{}}Counterpart in\\ Soft. Programming\end{tabular} & Speedup & Example \\ \hline
\hline
Explicit Data Caching     & Data Tiling 
& 5.6$\sim$32.1x                                                     & Fig.~\ref{fig:code}(a)    \\ \hline
Customized Pipelining     & \begin{tabular}[c]{@{}l@{}}Directive-Based\\ Programming\end{tabular}      & 1.3$\sim$10.3x                                                      & Fig.~\ref{fig:code}(b)     \\ \hline
PE Duplication            & Multithreading                                                                   & 1.0$\sim$53.6x                                                     & Fig.~\ref{fig:code}(b)     \\ \hline
Double Buffering          & \begin{tabular}[c]{@{}l@{}}Comp./Comm.\\ Overlapping\end{tabular}          & 1.0$\sim$2.1x                                                      & Fig.~\ref{fig:code}(c)     \\ \hline
Scratchpad Reorganization & Bit Packing
& 1.1$\sim$19.1x                                                      & Fig.~\ref{fig:code}(d)     \\ \hline
\end{tabular}
}
\end{table}

In summary, this paper makes the following contributions:




\begin{itemize}
\item A demonstration that the best-effort FPGA programming can achieve reasonable performance improvement (34.4x average speedup over a Xeon CPU core) with affordable programming efforts (as easy or difficult as CPU programming).
\item A best-effort guideline for mainstream software programmers to produce efficient FPGA accelerators, where open source code examples and quantitative performance evaluation of five major HLS optimizations (in Table~\ref{tab:strategy}) are presented in the iterative refinement. This also provides insights into source-to-source code transformations for compiler developers.
\end{itemize}


\section{Experimental Setup}
\label{sec:expsetup}

In this section we first provide a brief overview of modern CPU-FPGA platforms and then present the experimental setup used throughout this paper.

\subsection{CPU-FPGA Platform}

In CPU-FPGA platforms, FPGA compliments CPU cores by accelerating compute-intensive code regions 
with manageable offloading overhead, as summarized in \cite{cpu-fpga-dac16}. 

Today's mainstream PCIe-based FPGA boards can provide notable performance/watt improvement to datacenters~\cite{msft-catapult-isca14,yeung08,axel,fpmr,blaze} while requiring little system modification. To interface with the PCIe bus and on-board DRAM, a PCIe-DMA and a memory controller IP are required for implemention on the FPGA, in addition to user-defined accelerator logic. Fortunately, vendors have provided IP solutions to facilitate the design process. For example, Xilinx releases device support for the Alpha Data card in the SDAccel development environment~\cite{sdaccel}. Using such tools, users can focus on designing acceleration kernels and easily swap them into the device support to build customized FPGA accelerators.

Intel HARP integrates FPGA onto the same processor package. Through a QPI-based interconnect between the CPU and the FPGA, FPGA enjoys the benefit of having low-latency accesses to the host memory and a local coherent data cache. Moreover, virtual-to-physical address translation is supported to reduce the system software overhead and improve the host application programmability. However, users still need to undertake the same design process to program the FPGA, only interfacing with different peripheral IPs. 

In this paper we focus on the currently more accessible PCIe-based CPU-FPGA platform and HLS design flow to demonstrate the proposed best-effort guideline. Table~\ref{tab:exp_setup} lists the detailed hardware and software configuration. A Xeon CPU is connected with an Xilinx Virtex-7 FPGA board through the PCIe interface. For a fair comparison, both the CPU and the FPGA fabric were launched in 2012. On top of the platform hardware, we use Xilinx SDAccel to provide a hardware-software co-design environment.

\subsection{Benchmarks}
\label{subsec:bench}

This paper presents the proposed best-effort guideline through a complete accelerator design demonstration on a collection of benchmarks in MachSuite~\cite{machsuite}.
MachSuite is a benchmark suite that contains a broad class of computational kernels programmed as C functions for accelerator study.
For each kernel, MachSuite provides at least one implementation that is based on a commonly used algorithm in software programming, e.g., the queue-based algorithm for the BFS (breadth-first search) kernel.
This feature makes MachSuite a natural fit for demonstrating the proposed programmer-oriented guideline that aims to facilitate software programmers in refining pure software programs into high-quality FPGA accelerator designs.
Starting from the accelerators synthesized directly from the MachSuite kernel functions, Section \ref{sec:caching}, \ref{sec:frequency} and \ref{sec:communication} present the iterative accelerator refinement process.
Table \ref{tab:bench} lists all the kernels, each with a brief description.

The acceleration for the SORT (merge sort) kernel is relatively different from that of the others.
Merge sort has a tree-reduce characteristic, which means that the degree of parallelism will decrease by 2x after each merge layer.
The last few layers have very limited parallelism and are hard to be accelerated by FPGAs which heavily rely on parallelism to outperform CPUs. 
A common practice to resolve this issue is to let the FPGA accelerator focus on the first few layers and let the CPU do the remainder.
This paper adopts this approach and sets the goal of the SORT kernel to making every 1MB data chunk sorted.

\begin{table}[t]
\centering
\caption{Configuration of hardware and software.}
\label{tab:exp_setup}
{\scriptsize
\begin{tabular}{|l|l|}
\hline
Host CPU Model & Intel Xeon E5-2420 @ 1.9GHz (released in 2012) \\ \hline
Host Memory & 64GB DDR3-1600 \\ \hline \hline
FPGA Fabric & Xilinx Virtex-7 @ 200MHz (released in 2012) \\ \hline
Device Memory & 16GB DDR3-1600 (Max Band.: 12.8GB/s) \\ \hline \hline
CPU-FPGA Interface & PCIe Gen3 x8 (Max Band.: 8GB/s) \\ \hline
Synthesis Environment & SDAccel 2015.4 \\ \hline
\end{tabular}
}
\end{table}

\begin{table} [t]
  \centering
  \caption{Description of benchmarks used in the paper.}
  {\scriptsize
  \begin{tabular}
    {|l|l|} 
    \hline
    {Kernel}&{Description and Input Info.}\\
    \hline \hline
    AES     & \begin{tabular}[c]{@{}l@{}}Advanced encryption standard\\ Input: 256-bit key; 64MB data\end{tabular}                                                        \\ \hline
BFS     & \begin{tabular}[c]{@{}l@{}}Breadth-first search (queue-based)\\ Input: 4K nodes; 64K edges\end{tabular}                                        \\ \hline
GEMM    & \begin{tabular}[c]{@{}l@{}}General matrix multiplication ($N^{3}$ algorithm)\\ Input: two 1024x1024 double-precision matrices\end{tabular} \\ \hline
KMP     & \begin{tabular}[c]{@{}l@{}}Knuth-Morris-Pratt string matching\\ Input: 128MB string; 16B substring\end{tabular}                                \\ \hline
NW      & \begin{tabular}[c]{@{}l@{}}Needleman-Wunsch sequence alignment\\ Input: 64K pairs of 128-nucleotide sequences\end{tabular}                     \\ \hline
SORT    & \begin{tabular}[c]{@{}l@{}}Merge sort\\ Input: 64MB integer array\end{tabular}                                                                 \\ \hline
SPMV    & \begin{tabular}[c]{@{}l@{}}Sparse matrix-vector multiplication\\ Input: 4096x512 ELLPACK data and index matrices\end{tabular}                  \\ \hline
VITERBI & \begin{tabular}[c]{@{}l@{}}Viterbi algorithm\\ Input: 1M 128-element chains\end{tabular}                                                       \\ \hline
  \end{tabular}
  }
\label{tab:bench}
\end{table}

\section{Iter \#1: Explicit Data Caching}
\label{sec:caching}



We start from the naive FPGA accelerators directly synthesized from the original MachSuite kernel functions, which are slower than the Xeon CPU by 70$\sim$765x, as shown in Figure~\ref{fig:speedup-overall}.
Nonetheless, programmers are able to improve the performance of the accelerators by 5.6$\sim$32.1x by simply adding 15$\sim$20 lines of code to realize explicit data caching.
Section \ref{subsec:cache_pinpoint} pinpoints the performance bottleneck in the naive baseline and analyzes the underlying reason. 
Section \ref{subsec:cache_solution} presents the use of explicit data caching to resolve this.

\begin{figure}[!t]
\begin{minipage}{5cm}
\begin{lstlisting}[basicstyle={\scriptsize\ttfamily}]
void aes(...) { ... }

void kernel(char* data, int size) {
#pragma HLS interface m_axi port=data

  for (int i=0; i<size; i++) {
    aes(data+i*16);
  }
}
\end{lstlisting}    	
\end{minipage}
\begin{minipage}{3cm}
\raggedright
\includegraphics[scale=0.25]{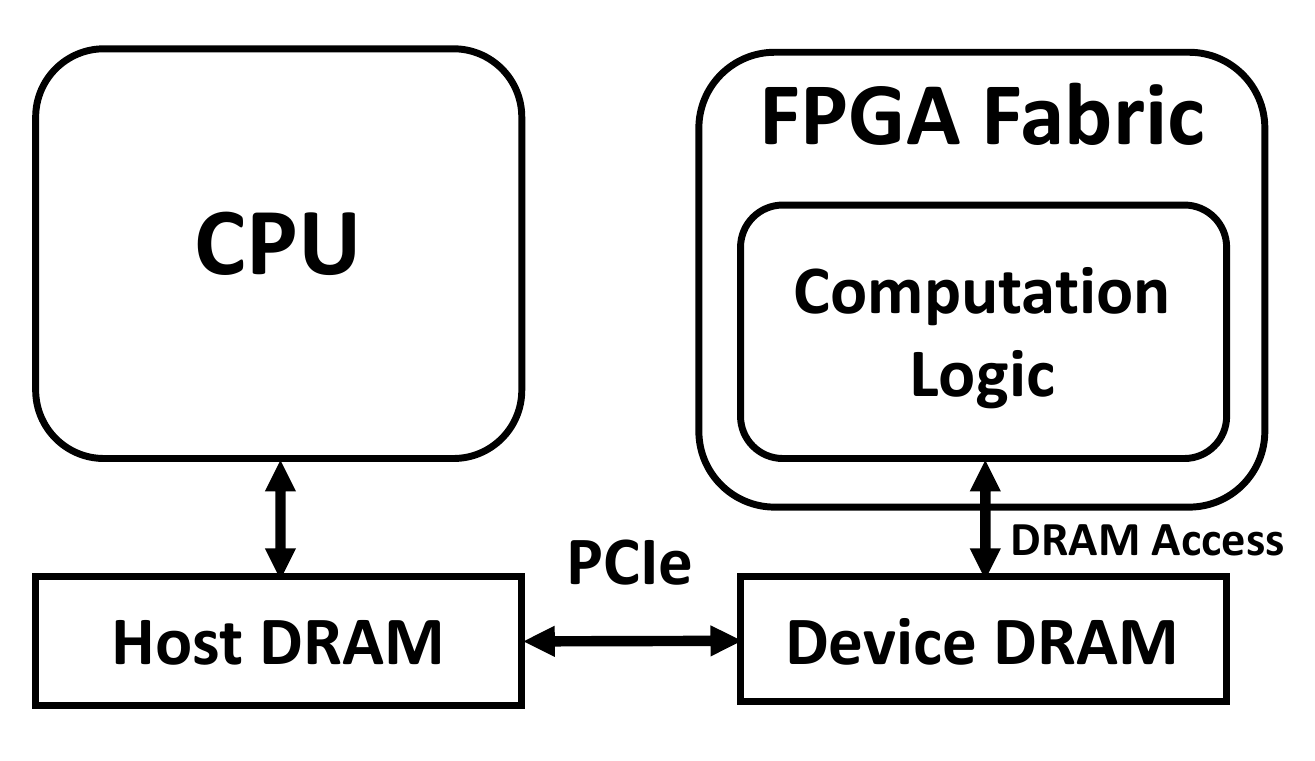}
\end{minipage}
\caption{Example AES kernel and naively generated architecture.}
\label{fig:baseline}
\end{figure}

To better demonstrate the refinement steps, we use the AES (advanced encryption standard) kernel as an example to deliver the code implementation of each step.
Figure~\ref{fig:baseline} shows the baseline AES kernel code\footnote{We ignore the key in this example to simplify the description. The complete AES kernel is available on our public repository.}.
The \textit{kernel} function accepts a certain \textbf{size} of \textbf{data}, and iteratively calls \textit{aes} function to encrypt the \textbf{data}.
Each \textit{aes} function call encrypts a 128-bit data block, so the \textbf{data} pointer shifts by 16 bytes after each iteration.
The \textit{interface} pragmas in the \textit{kernel} function specify the interface between the host program and the accelerator kernel.
The \textbf{data} are transferred from the host to the device through PCIe, and stored in the device DRAM.
We simply port the kernel to Xilinx SDAccel, with all implementation details of the \textit{aes} function unchanged.



\begin{figure}[!t]
  \centering
  \includegraphics[width=0.9\columnwidth]{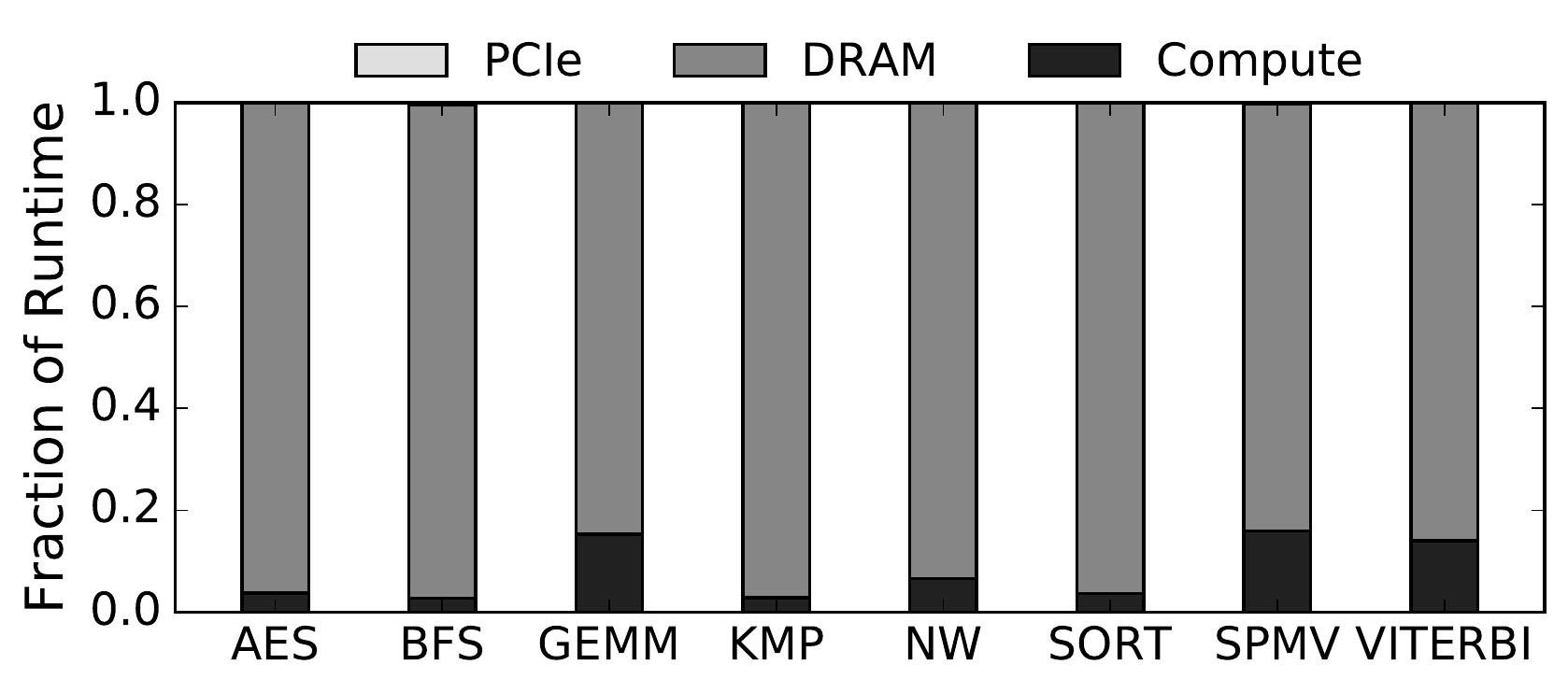}
  \caption{Execution time breakdown before any refinement.}
  \label{fig:breakdown-baseline}
\end{figure}


\begin{figure*}
\centering
\begin{minipage}[t]{\columnwidth}
\centering
\begin{lstlisting}[frame=t,belowskip=0pt]
void aes(...) { ... }

void load(char *buf, char *in) {
  memcpy(buf, in, BATCH_SIZE);
}

void store(char *out, char *buf) {
  memcpy(out, buf, BATCH_SIZE);
}

void compute(char *buf_data) {
  for (int i=0; i<BATCH_SIZE; i+=16) {
    aes(buf_data+i*16);
  }
}

void kernel(char *data, int size) {
  char buf_data[BATCH_SIZE];
  int batch_num = size/BATCH_SIZE;
  for (int i=0; i<batch_num; i++) {
\end{lstlisting}
\begin{lstlisting}[frame=none,backgroundcolor=\color{bg},aboveskip=0pt,belowskip=0pt]
    load(buf_data, data+i*BATCH_SIZE);
    compute(buf_data);
    store(data+i*BATCH_SIZE, buf_data);
\end{lstlisting}
\begin{lstlisting}[frame=b,aboveskip=0pt,belowskip=4pt]
  }
}
\end{lstlisting}
{\footnotesize \textbf{(a)} Applying explicit data caching}
\begin{lstlisting}[frame=t,belowskip=0pt]
int PE_BATCH = BATCH_SIZE / PE_NUM;

void aes(char *data) {
  for (...i...) {
\end{lstlisting}
\begin{lstlisting}[frame=none,backgroundcolor=\color{bg},aboveskip=0pt,belowskip=0pt]
#pragma HLS pipeline
\end{lstlisting}
\begin{lstlisting}[frame=none,aboveskip=0pt,belowskip=0pt]
  }
void load(...) { ... }
void store(...) { ... }

void compute(char *buf_data) {
  for (int j=0; j<PE_NUM; j++) {
\end{lstlisting}
\begin{lstlisting}[frame=none,backgroundcolor=\color{bg},aboveskip=0pt,belowskip=0pt]
#pragma HLS unroll
\end{lstlisting}
\begin{lstlisting}[frame=none,aboveskip=0pt,belowskip=0pt]
    for (int i=0; i<PE_BATCH ; i+=16)
      aes(buf_data[j]+i*16);
  }
}
void kernel(char *data, int size) {
  char buf_data[PE_NUM][PE_BATCH];
\end{lstlisting}
\begin{lstlisting}[frame=none,backgroundcolor=\color{bg},aboveskip=0pt,belowskip=0pt]
#pragma HLS array_partition var=buf_data cyclic=PE_NUM dim=1
\end{lstlisting}
\begin{lstlisting}[frame=b,aboveskip=0pt,belowskip=4pt]
  ...
}
\end{lstlisting}
{\footnotesize \textbf{(b)} Applying customized pipelining and PE duplication}
\end{minipage} \quad \quad
\begin{minipage}[t]{\columnwidth}
\centering
\begin{lstlisting}[frame=t,belowskip=0pt]
void aes(...) { ... }
void load(...) { ... }
void store(...) { ... }
void compute(...) { ... }
void kernel(char *data, int size) {
  char buf_data[3][PE_NUM][PE_BATCH];
#pragma HLS array_partition var=buf_data complete dim=1
#pragma HLS array_partition var=buf_data cyclic=PE_NUM dim=2

  for (int i=0; i < size/BATCH_SIZE; i++) {
\end{lstlisting}
\begin{lstlisting}[frame=none,backgroundcolor=\color{bg},aboveskip=0pt,belowskip=0pt]
    switch (i % 3) {
      case 0:
        load(buf_data[0], data+i*BATCH_SIZE);
        compute(buf_data[1]);
        store(data+i*BATCH_SIZE, buf_data[2]);
        break;
      case 1:
        load(buf_data[1], data+i*BATCH_SIZE);
        compute(buf_data[2]);
        store(data+i*BATCH_SIZE, buf_data[0]);
        break;
      case 2:
        load(buf_data[2], data+i*BATCH_SIZE);
        compute(buf_data[0]);
        store(data+i*BATCH_SIZE, buf_data[1]);
        break;
    }
\end{lstlisting}
\begin{lstlisting}[frame=b,aboveskip=0pt,belowskip=4pt]
  }
}
\end{lstlisting}
{\footnotesize \textbf{(c)} Applying double buffering}
\begin{lstlisting}[frame=t,belowskip=0pt]
void aes(...) { ... }
void load(...) { ... }
void store(...) { ... }
void compute(ap_uint<W> large_buf[][PE_BATCH]) {
  char normal_buf[BATCH_SIZE];
#pragma HLS array_partition var=normal_buf cyclic=PE_NUM dim=1
  for (int j=0; j<PE_NUM; j++) {
#pragma HLS unroll
\end{lstlisting}
\begin{lstlisting}[frame=none,backgroundcolor=\color{bg},aboveskip=0pt,belowskip=0pt]
    memcpy(normal_buf+j*PE_BATCH, large_buf[j], PE_BATCH);
    ... // parallel compute
    memcpy(large_buf[j], normal_buf+j*PE_BATCH, PE_BATCH);
\end{lstlisting}
\begin{lstlisting}[frame=none,aboveskip=0pt,belowskip=0pt]
  }
}
void kernel(ap_uint<W> *data, int size) {
\end{lstlisting}
\begin{lstlisting}[frame=none,backgroundcolor=\color{bg},aboveskip=0pt,belowskip=0pt]
  ap_uint<W> buf_data[3][PE_NUM][PE_BATCH];
\end{lstlisting}
\begin{lstlisting}[frame=b,aboveskip=0pt,belowskip=4pt]
  ...
}
\end{lstlisting}
{\footnotesize \textbf{(d)} Applying scratchpad reorganization}
\end{minipage}
\vspace{4pt}
\caption{Step-by-step example of applying major HLS optimization strategies to the AES benchmark. Major code rewriting regions involved in each step are highlighted.}
\label{fig:code}
\end{figure*}

\begin{figure*}
\centering
\subcaptionbox{}{
\includegraphics[height=3.2cm]{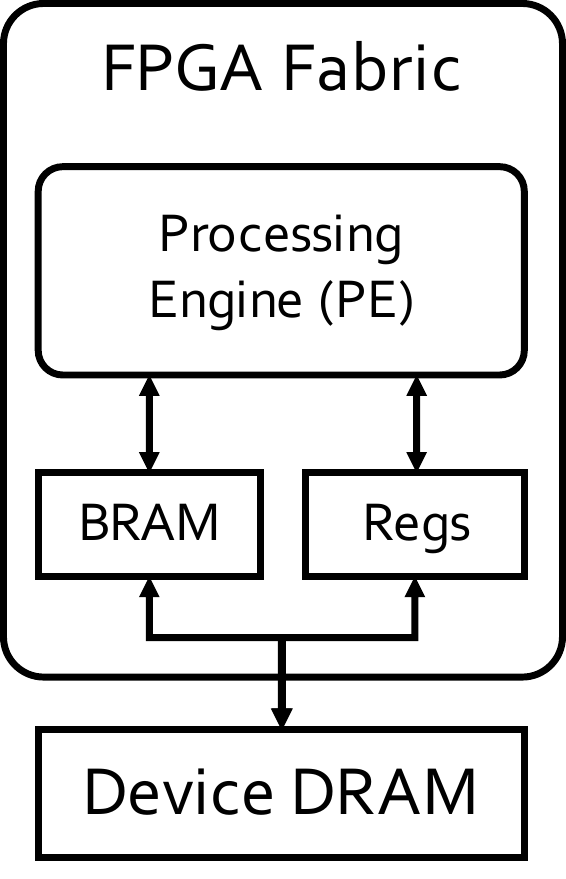}}
\enskip
\subcaptionbox{}{
\includegraphics[height=3.2cm]{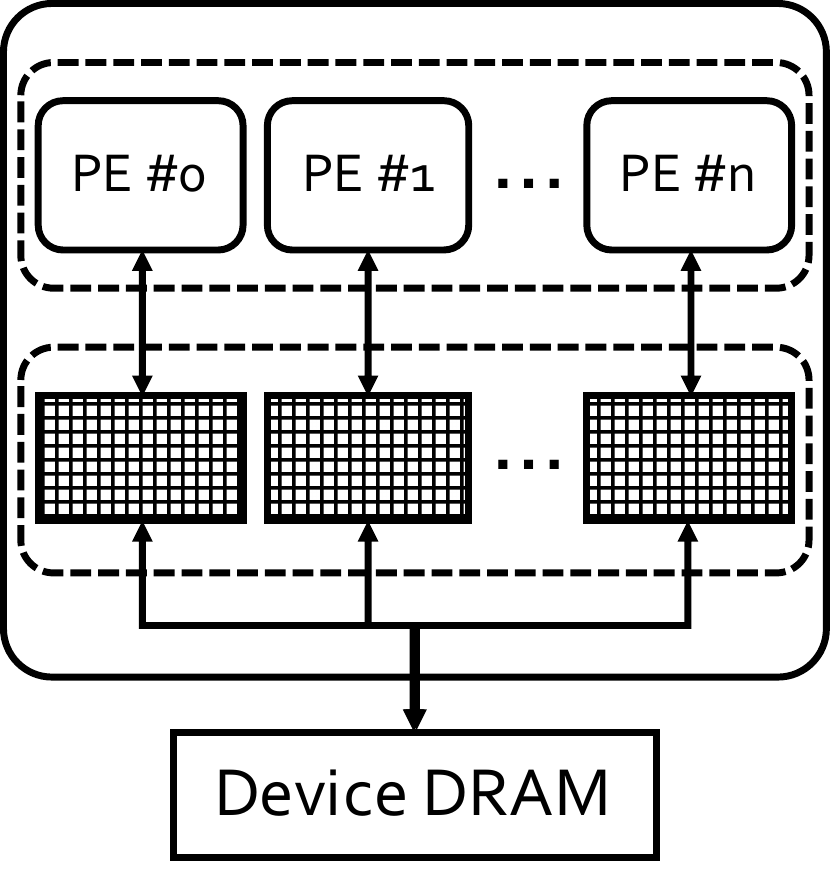}}
\enskip
\subcaptionbox{}{
\includegraphics[height=3.2cm]{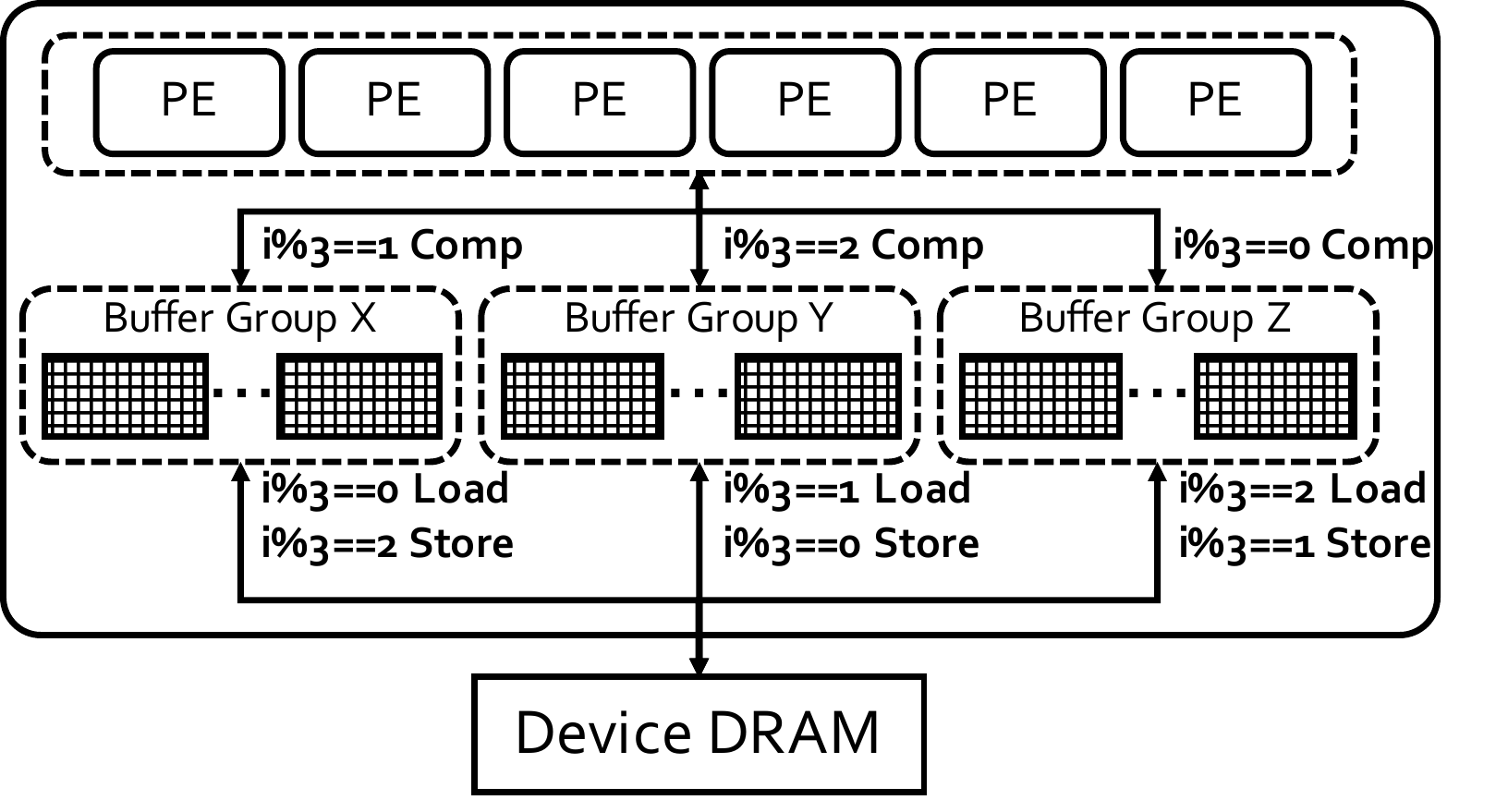}}
\subcaptionbox{}{
\includegraphics[height=3.2cm]{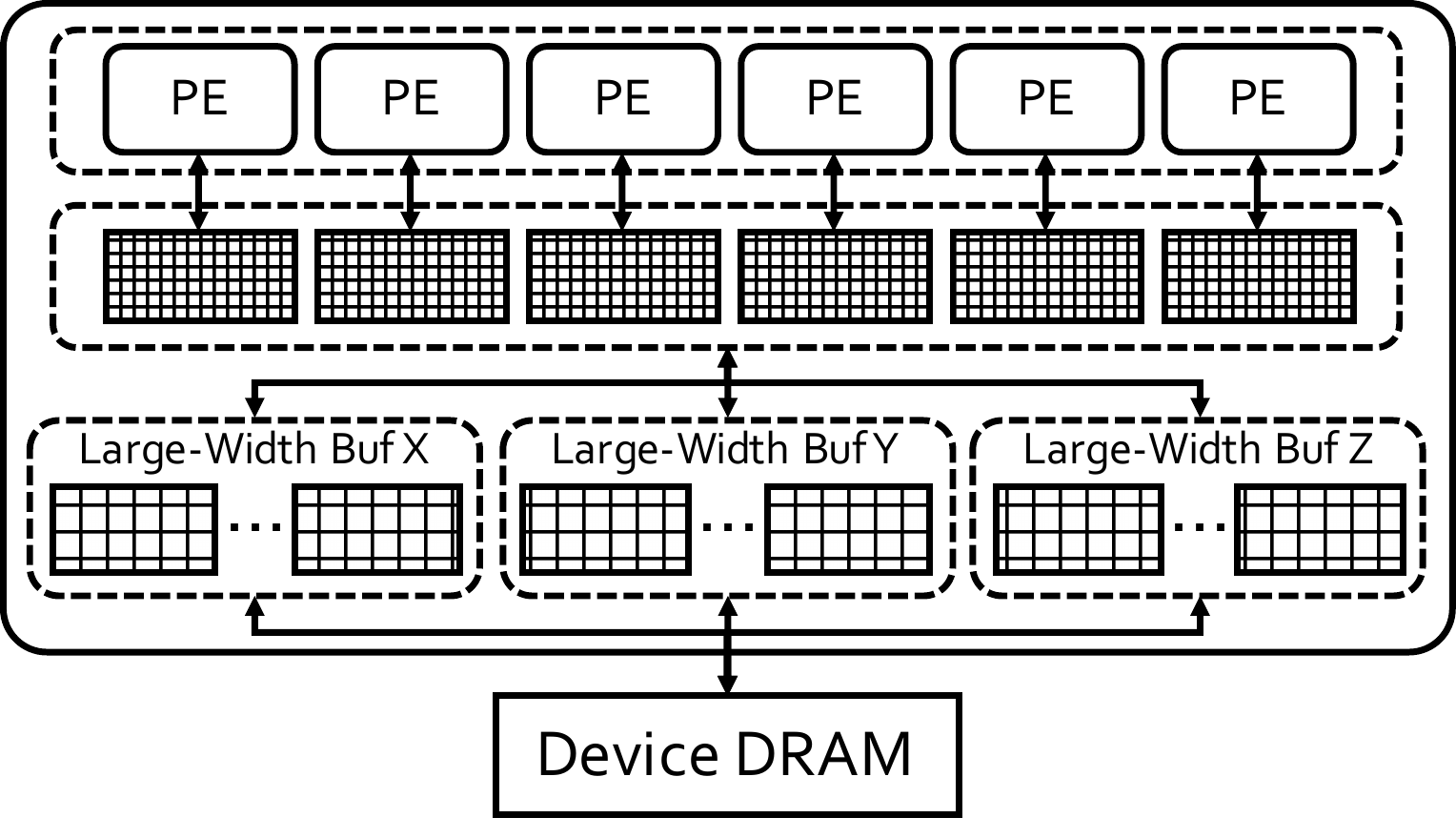}}
\caption{High-level architecture diagram of applying (a) explicit data caching, (b) PE duplication, (c) double buffering and (d) scratchpad reorganization, corresponding to Figure~\ref{fig:code}.}
\label{fig:arch}
\end{figure*}


\subsection{Cache: Not a Free Lunch Any More}
\label{subsec:cache_pinpoint}

Figure \ref{fig:breakdown-baseline} presents the execution time breakdown of the FPGA accelerators for the MachSuite kernels before any refinement.
It suggests that the DRAM access is dominating the overall execution for every kernel.
This is due to the fact that the cache hierarchy in CPUs, which provides a memory subsystem with low access latency while retaining programmer transparency, does not exist on FPGAs.
In contrast, FPGAs' on-chip BRAMs (block RAMs) that serve as the counterpart to caches are conceptually scratchpads, and have to be explicitly manipulated by software programs to realize data caching.
With such manipulation missed in the kernel function, the generated accelerator will connect the computation logic directly with DRAM with no high-speed data caching component in-between, as illustrated in Figure \ref{fig:baseline}.
Every data access has to physically go off chip, which costs a per-access initialization overhead of approximately 100 FPGA cycles (i.e., 500ns).

In summary, the programmer-transparent cache memory system is not a free lunch any more for HLS-based FPGA accelerator programming.  
In order to continue harnessing data caching to alleviate the DRAM access overhead, programmers must add code to explicitly cache data into the FPGA on-chip memory.

\subsection{Batch Processing and Data Tiling}
\label{subsec:cache_solution}

We present two techniques to implement explicit data caching in this paper.
One is batch processing which batches a number of jobs together and processes them in one action.
This approach is used for computational kernels whose working set sizes are far less than the total size of on-chip BRAM.
We use the AES kernel as an example to explain this approach. 
As introduced in the beginning of Section \ref{sec:caching}, an AES job, i.e., an \textit{aes} function call, encrypts only a 128-bit data block.
Since the working set size of an AES job is much smaller than the size of on-chip BRAM (a few MBs), programmers can maximize data reuse, i.e., temporal locality, by caching and processing one 128-bit data block at a time.
However, fetching 128-bit data at a time still leads to a serious DRAM access overhead because of the 100-cycle per-access initialization overhead.
An alternative approach is to process multiple contiguous 128-bit data encryption jobs in one batch.
With batch processing, multiple DRAM data fetches are combined into one memory burst operation, which spends 100 cycles in initialization and approximately 1 cycle (5ns) in fetching each piece of data.
The DRAM access overhead is then amortized.

The other technique is data tiling that first divides a job into a set of subjobs and then processes one or a few subjobs at a time.
This approach is used for computational kernels with relatively large working set sizes that are close to or far larger than the total size of on-chip BRAM.
We use the GEMM (general matrix multiplication) kernel as an example to explain the approach.
The GEMM kernel calculates the product of two matrices. 
While the matrices may be too large to be fully cached in BRAM, the matrix multiplication can be divided into additions and multiplications of submatrices, each of which can be as small as 1x1.
Programmers can then process one or more subjobs at a time to explore the temporal locality in the subjob level.

An important design choice is the caching size. 
The experimental platform used in this paper supplies approximately 4MB BRAM for FPGA accelerators (and the other few MBs for system-level IPs).
While a larger caching size that enables larger memory burst length is always beneficial in amortizing the 100-cycle initialization overhead, the effect of this amortization diminishes as the burst length increases.
Theoretically, if the payload size of a memory burst reaches 64KB, the burst length will be over 1000, and the impact of the initialization overhead will be reduced to less than 10\%.
Previous work also shows that the DRAM bandwidth can be saturated when the payload size comes to tens of KBs \cite{cpu-fpga-dac16}, which is less than 2\% of the total on-chip BRAM size provided by the FPGA fabric.
This indicates that explicit data caching can be realized with a very small BRAM consumption.

With these concepts in mind, the programming effort of implementing explicit data caching can be no more than that of writing 15$\sim$20 lines of code.
Figure~\ref{fig:code}(a) illustrates the implementation of explicit data caching through a code update of the AES baseline.
We can see that the overall execution of the AES kernel is decoupled into a series of \textit{load}-\textit{compute}-\textit{store} iterations.
Programmers only need to declare local arrays that represent on-chip BRAM buffers, and iteratively \textit{load} input the data of a batch of jobs (batch processing) or one or few subjobs (data tiling) to the arrays to \textit{compute}, and \textit{store} output data back to DRAM. 
The ``memcpy'' operations for loading/storing data will be inferred into memory read/write bursts to reduce the average DRAM access latency. 
This refinement leads to the addition of an intermediate BRAM layer between the computation logic and DRAM, as illustrated in Figure~\ref{fig:arch}(a).
Instead of letting the computation logic retrieve data directly from DRAM, this on-chip BRAM layer caches all necessary data for each iteration of computation.

\begin{figure}[t]
  \centering
  \includegraphics[width=\columnwidth]{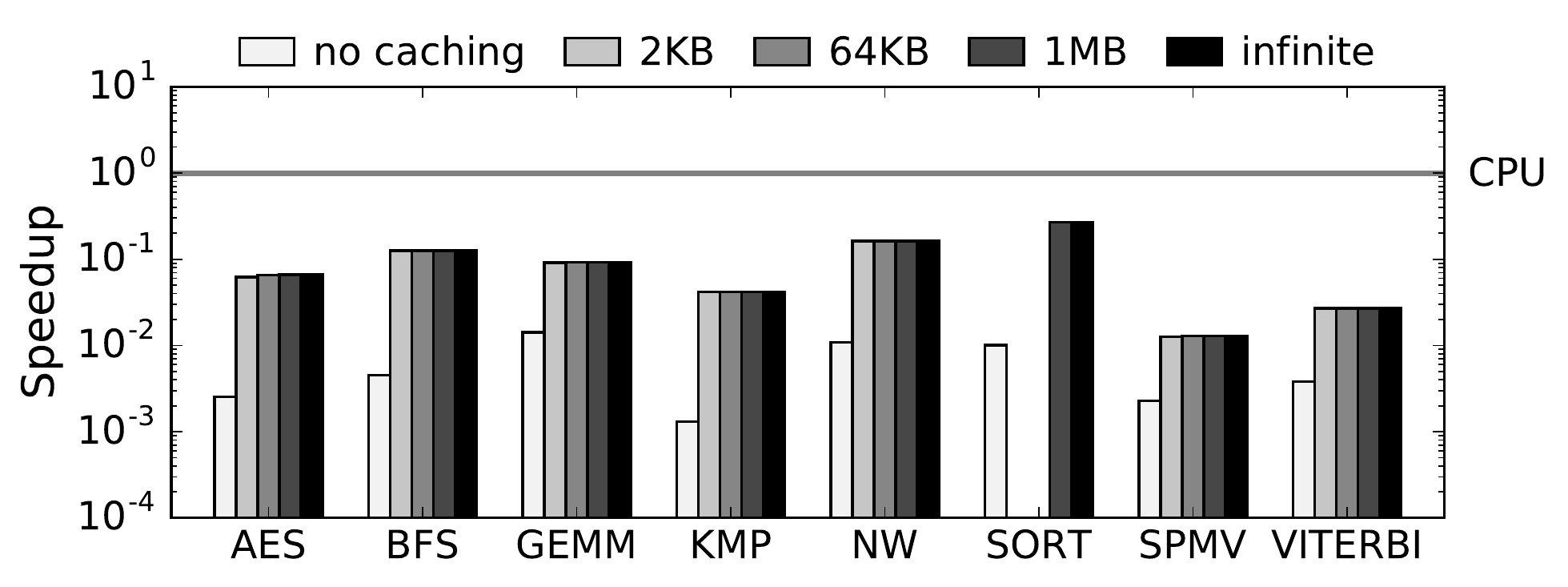}
  \caption{Normalized speedups in different caching sizes.}
  \label{fig:speedup-caching}
\end{figure}

Figure \ref{fig:speedup-caching} shows the normalized speedups of the accelerators compared to the Xeon CPU core after applying explicit data caching.
It also compares the performances of the accelerators with various caching sizes.~\footnote{The SORT kernel targets the sorting of each 1MB data chunk, so the caching size is set at 1MB only.}  
Each ``infinite'' bar delivers a speedup estimation where the caching size is infinite, so that no initial overhead of memory bursts is counted.
Two insights are revealed from the data.
First, explicit data caching results in a significant performance improvement over the naive baseline.
Second, the caching size has a negligible performance impact.
On one hand, the performances between the 64KB, 1MB and ``infinite'' groups are almost identical, 
which is consistent with our previous analysis on the design choice of caching size.
On the other hand, although there might be some performance differences between the 64KB and 2KB (close to the size of one BRAM block) groups, we observe that after explicit data caching is applied, the performance is dominated by computation (see Section~\ref{sec:frequency}). Thus, the performances between the 2KB and 64KB groups are also very similar.
This suggests that programmers can always consider shrinking down the caching size from the maximum ($\sim$4MB) to 1MB or 64KB to spare the BRAM resources for other optimization strategies.

\section{Iter \#2: Go Parallel}
\label{sec:frequency}



\begin{figure}[t]
  \centering
  \includegraphics[width=0.9\columnwidth]{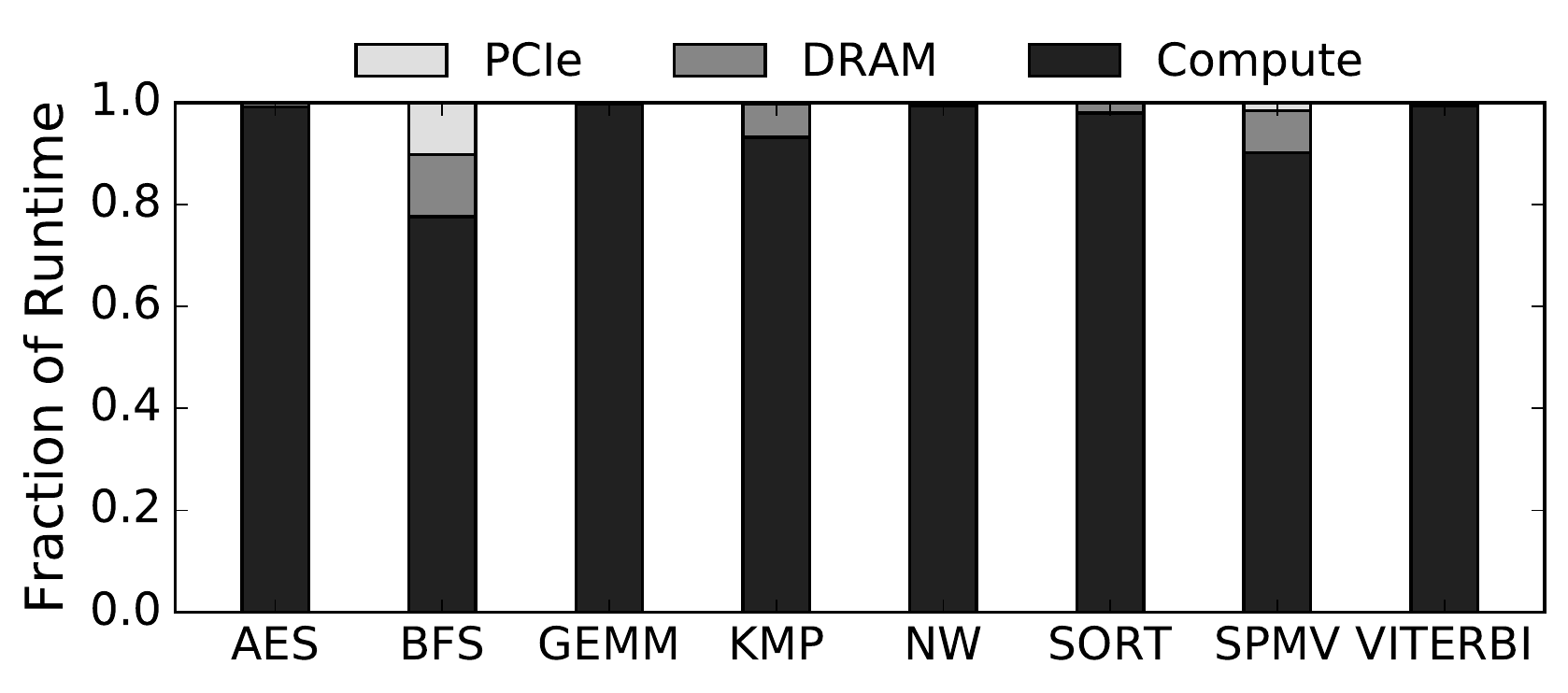}
  \caption{Execution time breakdown before Iteration \#2.}
  \label{fig:breakdown-iter1}
\end{figure}

Iteration \#2 starts from the accelerators that have applied explicit data caching.
%
Figure \ref{fig:breakdown-iter1} presents the execution time breakdown of the accelerators.
As the data hints, computation is dominating the overall execution for every kernel. 
The major reason is that the accelerators are doing computation sequentially with the culprit: a 200MHz clock frequency that is 9.5x lower than that of the Xeon CPU.
Consequently, FPGA accelerators heavily rely on exploring parallelism to dissolve the frequency disadvantage.
Section \ref{subsec:pipeline} and \ref{subsec:unroll} present the use of two principal parallelism exploration strategies to reduce the computation time: customized pipelining and processing element (PE) duplication.
With a programming effort of adding 10$\sim$15 lines of code, Iteration \#2 further improves the performance of the MachSuite accelerators by up to 417.8x.

\subsection{Iter \#2.1: Customized Pipelining}
\label{subsec:pipeline}
Pipelining is a fundamental concept in computer science and is used by both CPUs and accelerators.
Given no pipeline stalls, a pipeline with \textit{N} stages can boost the throughput by \textit{N} times.
Although various events, including branch misprediction and cache/TLB misses, impede the depth of a CPU pipeline to increase perpetually, 
FPGA accelerator designers can customize very deep pipelines for pure computational units with hundreds or even thousands of stages to greatly improve the accelerator performance.
Fine-grained pipelining to meet the maximum achievable data transfer speed almost becomes a de facto standard for FPGA accelerator designers.



Although customizing a full pipeline with an initiation interval (II) equal to 1 (i.e., the pipeline can process one iteration of data every cycle) would be difficult for mainstream software programmers, they can still harness the power of customized pipelining using a simplified approach.
In fact, it would be as easy as adding a sentence of ``pipeline'' pragma declaration to pipeline a loop block.
If a \textit{N}-iteration loop block with iteration latency \textit{L} is pipelined with initiation interval equal to \textit{ii} (II=\textit{ii}, and \textit{ii} is usually much smaller than \textit{L}), i.e., processing one loop iteration per \textit{ii} cycles, the execution time of the loop block can be reduced from \textit{N$\times$L} down to \textit{N$\times$ii+L}.
Since loops are often the most time-consuming code regions, this simple approach can actually lead to notable performance improvement.

\begin{table}[t]
\centering
\caption{Performance speedup of pipelining on computation.}
\label{tab:pipeline_res}
{\scriptsize
\begin{tabular}
    {|l c|l c|l c|} 
    \hline
    {Kernel}&{Speedup}&{Kernel}&{Speedup}&{Kernel}&{Speedup}\\
    \hline \hline
    {AES}&{1.4x}&{BFS}&{1.4x}&{GEMM}&{10.5x}\\
    \hline
    {KMP}&{7.0x}&{NW}&{8.8x}&{SORT}&{1.8x}\\
    \hline
    {SPMV}&{10.9x}&{VITERBI}&{3.2x}&&\\
    \hline
  \end{tabular}
}
\end{table}

We evaluate all 40 loop blocks in all benchmarks by simply adding the ``pipeline'' pragma inside each loop block's innermost loop.
The result shows that 27 loop blocks can be immediately pipelined, and 6 loop blocks can be pipelined if being transformed into perfect loops.
Moreover, a significant speedup on the computation is observed for many kernels, as listed in Table \ref{tab:pipeline_res}.
Some kernels such as SPMV, NW, GEMM and KMP reach a 7.0$\sim$10.9x speedup, because the main bodies of these kernels are nested loop blocks.
SPMV and GEMM do linear algebra in a two-level and three-level nested loop, respectively; NW does a two-dimension dynamic programming in a two-level nested loop; KMP matches a substring in a string in a two-level nested loop.
These loop bodies are well pipelined.
Other kernels such as AES, BFS and SORT have relatively complicated kernel function bodies, and reach relatively moderate speedup---40\%$\sim$80\% performance improvement.
VITERBI is a delicate case. Although it also does a dynamic programming in a nested loop body, it requires each pipeline stage to complete a few \emph{floating-point} additions, multiplications and comparisons (subtractions), which results in a pipeline with a relatively large II.
In contrast, the NW kernel, with a similar computation pattern, requires each pipeline stage to complete merely a few \emph{low-width integer} additions and \emph{bit-level} comparisons, which can be finished in one cycle, i.e., achieving an II=1 pipeline.
Therefore, the speedup for VITERBI (3.2x) is fairly less than that for NW (8.8x), but still considerable.

\subsection{Iter \#2.2: Processing Element Duplication}
\label{subsec:unroll}
Processing element (PE) duplication explores the task-level parallelism in kernels.
If a large number of independent jobs can be found in a kernel, then programmers can create multiple PEs to process them in parallel.
This concept is not new to software programmers.
With multicore becoming ubiquitous in modern processors, programmers have been accustomed to mapping independent jobs onto multiple cores and do them in parallel.
Here, FPGA PEs and CPU cores are counterparts.
As a consequence, the implementation of PE duplication can be considered as a special ``multithreading programming,''

Figure~\ref{fig:code}(b) illustrates the implementation of customized pipelining and PE duplication in one code example since the two strategies work on mutually exclusive code regions.
This code example is updated from the one in Figure~\ref{fig:code}(a), which implements explicit data caching.
We intentionally omit the implementation details that remain unchanged so as to highlight the newly added code regions.
This update features three major changes.
First, the ``pipeline'' pragma is added into each loop block of the \textit{aes} function to perform customized pipelining.
Second, the ``unroll'' pragma is adopted to generate multiple PE duplicates for parallel computation.
In addition, the ``memory partition'' pragma is used to partition the local arrays used for explicit data caching into multiple segments, the number of which is equal to the PE duplication factor.
The first change pipelines the loop blocks, and the latter two changes together realize PE duplication.
Figure~\ref{fig:arch}(b) illustrates the refined architecture after applying pipelining and PE duplication.
Compared to Figure~\ref{fig:arch}(a), the refined architecture partitions the intermediate BRAM layer into multiple groups, each of which communicates only with one PE duplicate.

\begin{figure}[t]
\centering
\includegraphics[width=0.5\columnwidth]{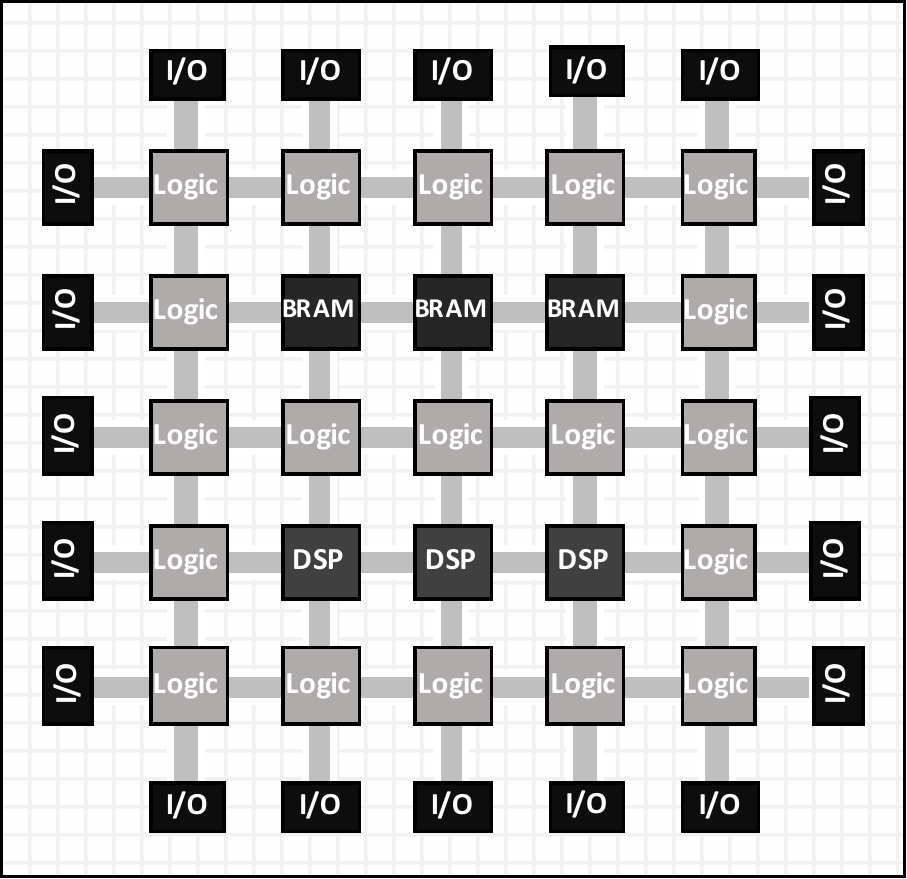}
\vspace{2pt}
\caption{Architecture of an FPGA fabric.}
\label{fig:fpga}
\end{figure}

One thing worthwhile mentioning here is memory partitioning, whose objective is to realize parallel data supply for all PE duplicates. 
FPGA is a kind of reconfigurable logic that contains distributed computation building blocks---LUTs (lookup tables) and DSPs (digital signal processors) and BRAM (distributed on-chip memory building blocks)---as illustrated in Figure \ref{fig:fpga}.
The computation building blocks enable the creation of multiple PE duplicates, and the BRAM blocks supply a many-port on-chip memory system. 
Specifically, the Xilinx Virtex-7 FPGA fabric has approximately 3000 BRAM blocks, i.e., a virtually 3000-port on-chip memory system which has the potential to feed up to 3000 PEs concurrently.
To fulfill this potential, however, a programmer needs to partition the local array, i.e., the on-chip BRAM buffer for data caching, into multiple segments, each  made up of a set of BRAM blocks.
When input data are loaded into the BRAM buffer, they will be scattered into different segments and processed by different PEs simultaneously. 

\begin{figure}[t]
  \centering
  \includegraphics[width=0.9\columnwidth]{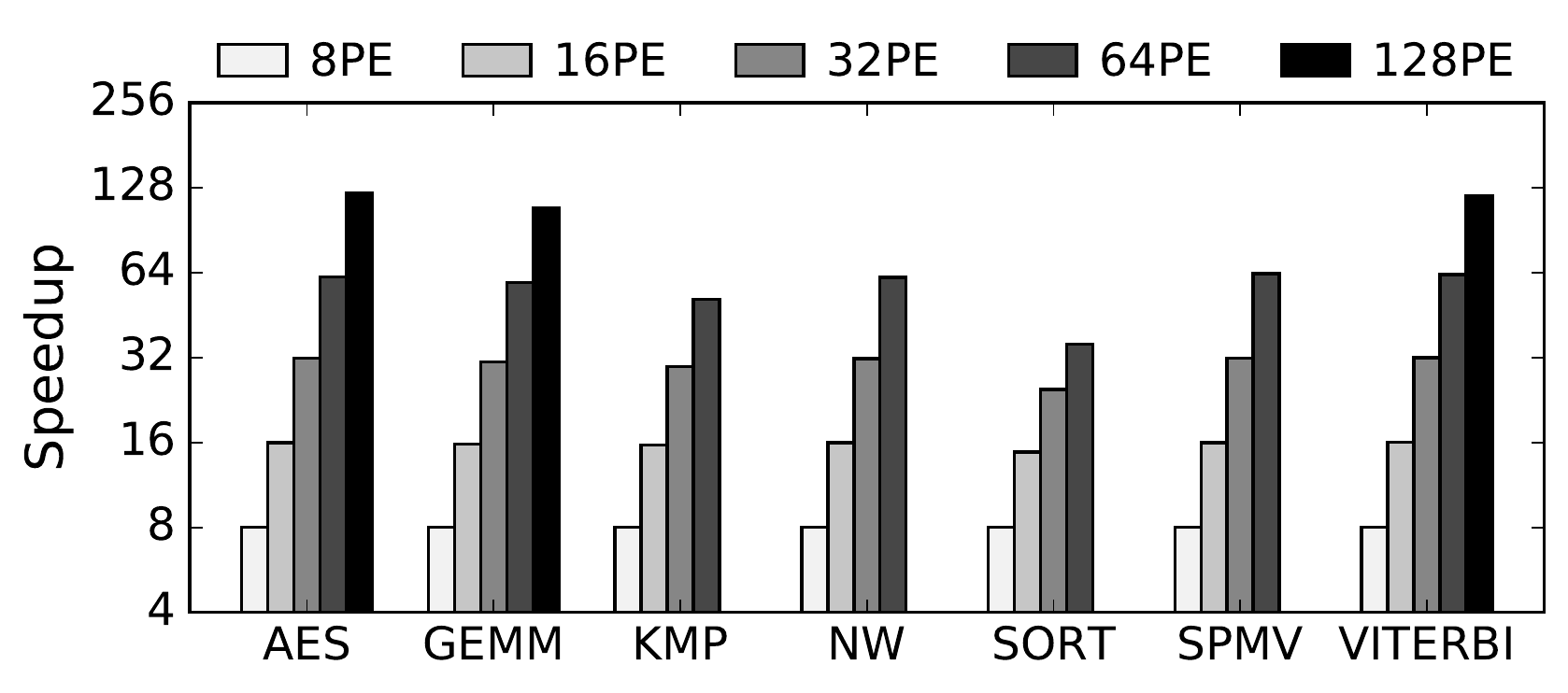}
  \caption{Performance speedup on computation lead by PE duplication, compared to the one PE baseline.}
  \label{fig:pe-duplication}
\end{figure}

Figure \ref{fig:pe-duplication} compares the performance improvement on computation with various PE duplication factors.\footnote{Some kernels may not generate a 128-PE design due to FPGA resource constraints. The corresponding bar is thus left blank.}
For each kernel, we normalized all the performances to that of the accelerator with one PE to ease the observation.
Most kernels achieve a linear performance improvement. Such kernels as AES, NW and VITERBI can be divided into fully parallel jobs and thus reach close-to-ideal speedup.
The performance of the SORT kernel does not scale linearly due to its tree-reduce characteristic, i.e., the degree of parallelism is reduced by 2x after each merge layer. Therefore, the last few merge layers will have less degrees of parallelism than the number of PEs available, and it cannot be accelerated in fully parallel.
For the BFS kernel whose jobs are chain-dependent because of the sequentially accessed queue structure, PE duplication is not applicable and thus the kernel is not shown in Figure \ref{fig:pe-duplication}.
%


\begin{figure}[t]
  \centering
  \includegraphics[width=\columnwidth]{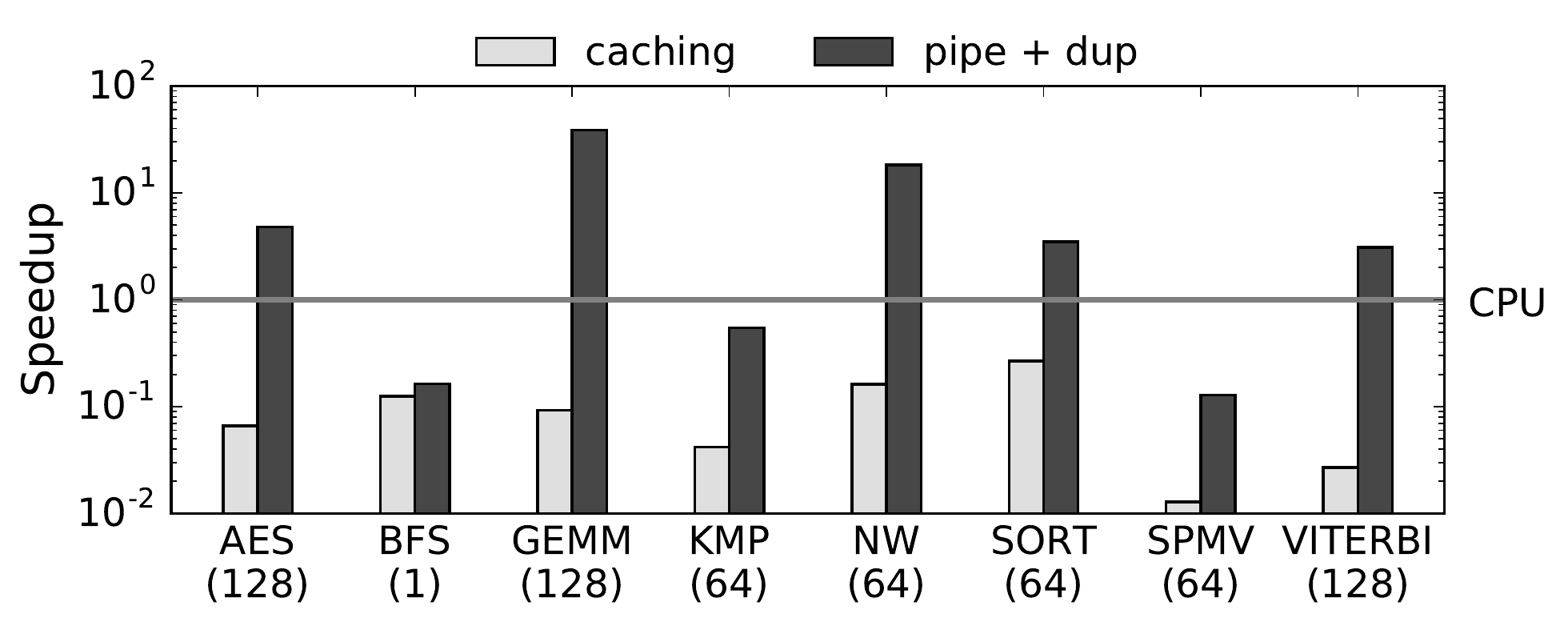}
  \caption{Overall performance speedup after applying local loop pipelining and PE duplication.}
  \label{fig:speedup-dup-pipe}
\end{figure}

\subsection{Overall Speedup}

Figure \ref{fig:speedup-dup-pipe} presents the overall speedup of each accelerator over the Xeon CPU core after implementing pipelining and PE duplication.
The number under each kernel's name represents the best PE duplication factor.
The horizontal line (1) represents the CPU baseline with the bars above the line representing speedup and below representing slowdown.
Most accelerator designs have two-orders-of-magnitude speedups over those in Iteration \#1 and start to outperform the CPU core.
But the speedups are still considerably far from satisfactory.
This is due to the fact that the DRAM access overhead comes back again to play an important role in the overall execution (see Figure \ref{fig:breakdown-iter2}) after the computation routine is significantly accelerated.
Section \ref{sec:communication} further optimizes the data movement to address this issue.



\section{Iter \#3: Faster Data Movement}
\label{sec:communication}



\begin{figure}[!t]
  \centering
  \includegraphics[width=0.9\columnwidth]{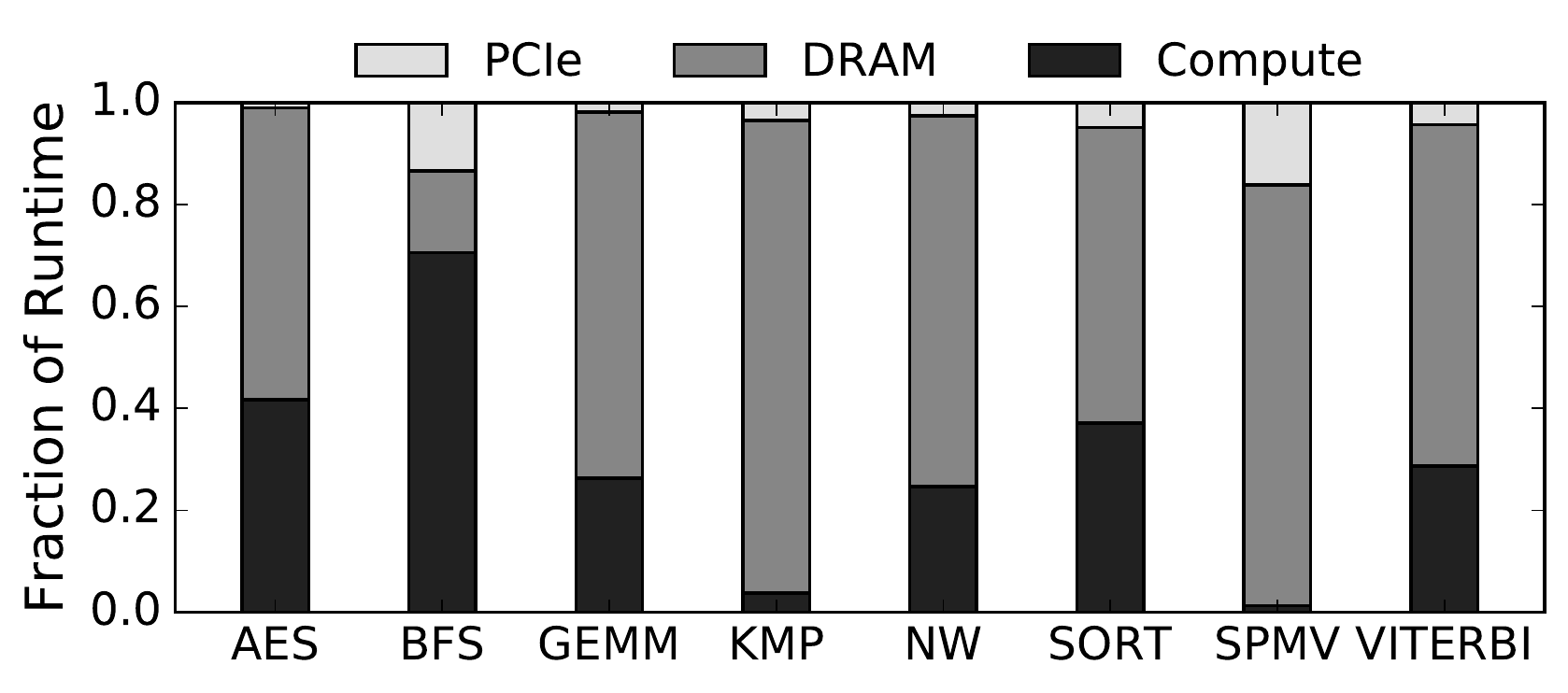}
  \caption{Execution time breakdown before Iteration \#3.}
  \label{fig:breakdown-iter2}
\end{figure}

Iteration \#3 starts from the accelerators that have applied explicit data caching, pipelining and PE duplication.
As shown in Figure \ref{fig:breakdown-iter2}, DRAM access becomes the major performance bottleneck again.
Section \ref{subsec:overlap} and Section \ref{subsec:utilization} present the use of double buffering and scratchpad reorganization to increase the DRAM bandwidth utilization from the temporal and spatial aspects, respectively. 
With a programming effort of adding around 100 lines of code, the performance of the MachSuite accelerators can be further improved by 1.2$\sim$19.2x.
Moreover, the code implementation of the two strategies is highly reusable across different kernels, so the programming effort can be amortized by multiple design practices.




\subsection{Double Buffering}
\label{subsec:overlap}

As is illustrated in Figure~\ref{fig:code}(b), although the accelerator designs perform computation in parallel, each PE processes different \textit{load}-\textit{compute}-\textit{store} iterations sequentially.
In other words, the \textit{N-th} iteration starts to load data after the \textit{(N-1)-th} iteration stores data back to DRAM.
However, the data loading of the \textit{N-th} iteration could have happened earlier, right after the \textit{(N-1)-th} iteration finishes loading data and starts computation, since the read channel of the AXI bus, which interfaces between the accelerator and DRAM, becomes free.
In general, the \textit{load}, \textit{compute} and \textit{store} procedures of adjacent iterations can be overlapped to form a 3-stage coarse-grained pipeline, which results in an improved resource utilization as well as a better performance.
The proposed best-effort guideline uses the double buffering strategy to realize such a coarse-grained pipeline.

Figure~\ref{fig:code}(c) illustrates the implementation of double buffering through an update of the code example in Figure ~\ref{fig:code}(b).
This update features two major changes.
First, the local arrays for explicit data caching are duplicated into three identical copies.
This corresponds to the architectural change where the intermediate BRAM layer is duplicated into three identical BRAM buffer groups, as illustrated in Figure~\ref{fig:arch}(c). 
Second, a switch/case statement is added to schedule the \textit{load}, \textit{compute} and \textit{store} procedures.
This schedule is the key to realizing double buffering.

To better explain the scheduling mechanism, we first make the following denotations.
We use the variable names in the code example to denote the three buffer groups, i.e., \textit{buf\_data[0]}, \textit{buf\_data[1]} and \textit{buf\_data[2]}, corresponding to the hardware component \textbf{X}, \textbf{Y} and \textbf{Z} in Figure~\ref{fig:arch}(c).
We also index the execution phases by letter \textbf{i}.
In addition, we denote the input and output data of the \textit{k-th} iteration as \textit{$I_{k}$} and \textit{$O_{k}$}.
The scheduling mechanism is then described as follows.

\begin{itemize}
\item \textit{\textbf{i == 0}}: load \textit{$I_{0}$} into \textit{buf\_data[0]} 
\item \textit{\textbf{i == 1}}: load \textit{$I_{1}$} into \textit{buf\_data[1]}; process \textit{$I_{0}$} in \textit{buf\_data[0]}
\item \textit{\textbf{i == 2}}: load \textit{$I_{2}$} into \textit{buf\_data[2]}; process \textit{$I_{1}$} in \textit{buf\_data[1]}; store \textit{$O_{0}$} and free \textit{buf\_data[0]}
\item \textit{\textbf{i == 3}}: load \textit{$I_{3}$} into \textit{buf\_data[0]}; process \textit{$I_{2}$} in \textit{buf\_data[2]}; store \textit{$O_{1}$} and free \textit{buf\_data[1]}
\item ...
\end{itemize}


As shown in Figure~\ref{fig:speedup-lcs-reorg}, double buffering contributes up to 2.1x performance improvement.
Most kernels achieve at least a 20\% performance improvement.
The BFS kernel cannot be benefited from this technique, mainly because the queue-based searching mechanism determines that the compute results of an iteration will affect the input data to load in the next iteration.
KMP is another kernel of which the performance is almost not changed, which is mainly due to the fact that the output of KMP is merely an integer representing the number of substrings found in MachSuite.

\subsection{Scratchpad Reorganization}
\label{subsec:utilization}

When it comes to the spatial aspect, the issue of DRAM bandwidth utilization becomes delicate.
We use a piece of C code to reveal the issue.
List \ref{code:demo} shows four C statements that declare four arrays. While defined in different types with different lengths, all such arrays represent a 1KB contiguous memory space and are equivalent from a CPU programmer's perspective.
Specifically, each type of array can be cast to and used as any other type, as shown in List \ref{code:demo}.

\begin{lstlisting}[caption=Sample code to demonstrate the difference between CPUs and FPGAs in interpreting arrays., label=code:demo]
char arr_byte[1024];      // Statement #1
short arr_short[512];     // Statement #2
int arr_int[256];         // Statement #3
long long arr_ll[128];    // Statement #4

// cast and use int array as char array
char* p = (char *)arr_int;
for (i=0; i<1024; i++) p[i] = 0;

// cast and use short array as long long array
long long* q = (long long *)arr_short;
for (i=0; i<128; i++) q[i] = 0;
\end{lstlisting}

In FPGA programming using HLS-C, however, the above four arrays are essentially different from each other and cannot be cast to and used as other types.
We use Statement \#1 and \#3 to explain the difference.
Statement \#1 defines a 1024-entry byte array, which is synthesized into an on-chip BRAM buffer with width 8 bits and depth 1024.
In other words, an accelerator can at most read a byte of data in each FPGA cycle from this buffer, and thus takes at least 1024 cycles to traverse it.
In contrast, Statement \#3 represents a BRAM buffer with width 32 bits and depth 256, indicating that it can be traversed in 256 FPGA cycles with each cycle reading 32-bit data.
In fact, SDAccel supports a BRAM buffer to have up to 512-bit data width, but primitive C types have at most 64-bit width.
As a result, the DRAM bandwidth utilization can achieve at most 12.5\% of the ideal value, and will be less than 2\% of the ideal value if the buffer is defined as the \textit{char} type. 

The above analysis makes it clear that programmers can increase the BRAM bandwidth utilization by declaring BRAM buffers used for explicit data caching with larger widths.
HLS-C provides a large-width integer type template - \textit{ap\_int<W>}, where \textit{W} denotes the width of the data type.
Harnessing this template, we propose the scratchpad reorganization technique to increase DRAM-BRAM transfer bandwidth.

Figure~\ref{fig:arch}(d) illustrates the proposed approach which makes two major changes on the accelerator architecture.
First, we replace the three BRAM buffer groups, \textbf{X, Y and Z} in Figure~\ref{fig:arch}(c), with three new BRAM buffer groups with the same capacity but larger width.
The value of the buffer width is restricted to be the power of two between 8 and 512, so as to be compatible with C types and the AXI bus width (512 bits) between the accelerator and DRAM.
Second, we add another BRAM layer which is identical to that in Figure~\ref{fig:arch}(b), i.e., one of the \textbf{X, Y and Z} in Figure~\ref{fig:arch}(c).
In other words, while the three buffer groups are updated with larger width, we still keep a copy of the original, normal-width buffer group to directly communicate with PEs.
Given these architectural changes, the \textit{load}, \textit{compute} and \textit{store} procedures are updated as follows.

\begin{itemize}
\item \textbf{\textit{Load.}} Loading input data from DRAM to one of the large-width BRAM buffer group
\item \textbf{\textit{Compute.}} \\
- Transferring input data from the large-width to normal-width BRAM buffer groups \\
- Parallel computing, which remains unchanged \\
- Transferring output data from the normal-wdith to large-width BRAM buffer groups
\item \textbf{\textit{Store.}} Storing output data from one of the large-width BRAM buffer group back to DRAM
\end{itemize}

The code update to implement this scratchpad reorganization strategy is illustrated in Figure~\ref{fig:code}(d).
This approach increases the computation time by adding two BRAM-BRAM data transfers in the \textit{compute} function. However, since the bottleneck is on the DRAM side, and both the large-width and normal-width buffer groups are partitioned, thus transferring data in parallel, the overhead on computation does not seriously harm the overall performance.

Resource constraints play an important role in applying scratchpad reorganization.
Given a certain capacity, a BRAM buffer with a larger width usually consumes more BRAM blocks.
Specifically, a BRAM block in the Virtex-7 fabric has a 18Kb capacity with at most 36-bit width. It costs at least 8 blocks to construct a 256-bit BRAM buffer, and 15 blocks for a 512-bit buffer.
For an accelerator design with 128 PEs, it costs at least 5760 BRAM blocks to allow 128 large-width BRAM buffers for all three buffer groups, while there are only around 3000 BRAM blocks available on the FPGA fabric.
However, programmers only need to try at most three design choices for each kernel to reach the best trade-off between the BRAM buffer width and the PE duplication factor (see Section \ref{sec:discussion}).

As shown in Figure \ref{fig:speedup-lcs-reorg}, for the scratchpad reorganization strategy, the KMP and AES kernels achieve significant speedups since their original input/output types are the 8-bit \textit{char} type.
A \textit{char}-type BRAM buffer can be enlarged to an \textit{int}-type buffer without even consuming any more BRAM, since a BRAM block can be configured into a up-to-36-bit buffer.
As a consequence, these two kernels can be greatly improved via scratchpad reorganization without consuming too much BRAM.
On the other hand, kernels such as SPMV and GEMM have already used wider C types, such as int, float and double.
Each increment of the buffer width may lead to up to 2x BRAM consumption, and the speedup is thus limited.

\subsection{Final Results}
\begin{figure}[!t]
  \centering
  \includegraphics[width=\columnwidth]{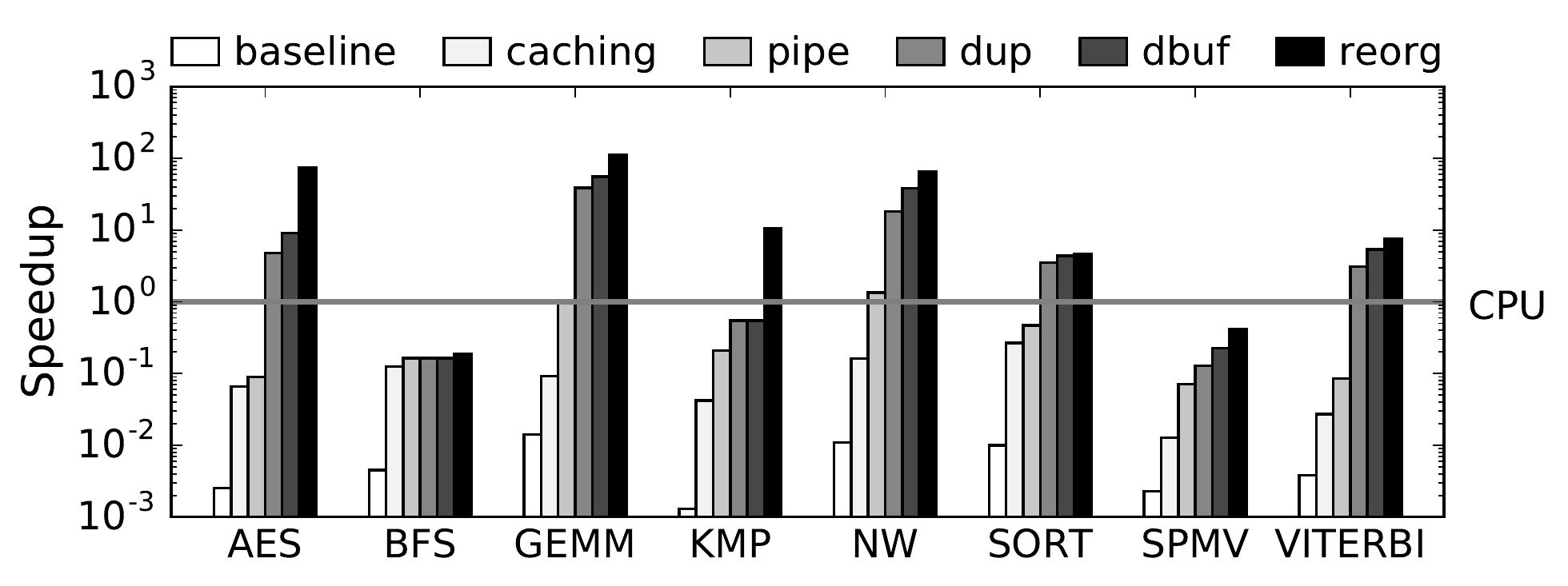}
  \caption{Performance improvement by applying five optimization techniques step by step (accumulative).}
  \label{fig:speedup-lcs-reorg}
\end{figure}

Iteration \#3 concludes the demonstration of the proposed best-effort guideline.
Figure \ref{fig:speedup-lcs-reorg} summarizes the performance improvement of each optimization strategy in the refinement steps.
Except for the SPMV and BFS kernels that have been determined non-acceleratable in our experimental platform even before any refinement iteration (see Table~\ref{tab:pcie} in Section \ref{sec:discussion}), all the kernels have outperformed CPU by at least 4.7x.
50\% of the kernels achieve at least an order-of-magnitude speedup.
Meanwhile, compared to the naive accelerators generated from the original software kernels in MachSuite, the proposed flow brings 42$\sim$29030x performance improvement, which demonstrates the effectiveness of our programmer-friendly iterative design guideline with five major HLS optimizations.
Moreover, as illustrated throughout this paper, these optimization techniques, which can be well understood by software programmers, require a fairly small amount of programming effort.
In fact, all of them can easily find software programming counterparts, such as data tiling, directive-based programming, multithreading, computation/communication overlapping and bit packing, as summarized in Table~\ref{tab:strategy}.


\section{Discussion}
\label{sec:discussion}

After the demonstration of the proposed practice guideline, we start to discuss its adaptability.
This section is organized in a Q\&A format.
We list a series of important questions and corresponding answers to assist programmers in learning when and how to use the flow in accelerator development.

\noindent \textbf{Q: Why do people think FPGA programming is difficult? Why can we make it easy?}

Conventionally people treat FPGA as a piece of specialized hardware to accelerate special-purpose applications. 
Therefore, they focus on achieving the best performance by applying a combination of optimization techniques (will be introduced in Section~\ref{sec:related}, including both software programmer accessible and inaccessible optimizations) from an exponential design space.
This often requires intimate hardware expertise.

With Intel's acquisition of Altera, the role of FPGAs is changing: it is becoming a mainstream computing resource and ``free lunch'' in future servers.
Therefore, the goal or expectation of FPGA accelerator design is also changing: the optimal performance is not always required; instead, mainstream software programmers want more accessible programming guidelines and reasonable performance improvement (after all, it is almost a free bonus).
With such a new role and expectation of FPGAs, we are the first to re-examine FPGAs' programming practice and find that following our proposed guideline, a few programmer-friendly optimization steps can produce compelling FPGA accelerators.

\noindent \textbf{Q: What role does the CPU-FPGA communication play in the acceleration process?}

While previous work often omits host-device communication when calculating speedup, this paper considers the system-level speedup that includes all the computation off-
loading overhead from a CPU to an FPGA, which is more practical in real deployment of FPGAs in servers.
In our platform that uses the PCIe connection between the CPU and FPGA, we calculate the PCIe transfer time as the elapsed time of data movement from the host memory to the device memory through PCIe-based direct memory access (DMA).
Although the optimization on PCIe transfer is beyond the scope of this paper, the PCIe transfer time (or more generally, the CPU-FPGA communication time) serves as a valuable indicator to filter out the FPGA acceleration for communication-bounded kernels before any refinement.

Table \ref{tab:pcie} lists the PCIe transfer time of the MachSuite kernels used in the paper.
Each kernel's PCIe transfer time is normalized to its execution time on the Xeon CPU. 
While most kernels have negligible PCIe transfer time and large speedup potentials, the BFS and SPMV kernels show severe PCIe transfer overheads.
This explains why the BFS and SPMV kernels are not accelerated in our experimental platform.

\begin{table} [h]
  \centering
  \caption{PCIe transfer time normalized to CPU runtime}
{\scriptsize
\begin{tabular}
    {|l c|l c|l c|} 
    \hline
    {Kernel}&{PCIe}&{Kernel}&{PCIe}&{Kernel}&{PCIe}\\
    \hline \hline
    {AES}&{$2.2 \times 10^{-3}$}&{\textbf{BFS}}&{\textbf{0.8}}&{GEMM}&{$6.0 \times 10^{-4}$}\\
    \hline
    {KMP}&{$5.9 \times 10^{-2}$}&{NW}&{$1.5 \times 10^{-3}$}&{SORT}&{$4.9 \times 10^{-3}$}\\
    \hline
    {\textbf{SPMV}}&{\textbf{1.3}}&{VITERBI}&{$1.4 \times 10^{-2}$}&&\\
    \hline
  \end{tabular}
}
\label{tab:pcie}
\end{table}


\noindent \textbf{Q: What role do FPGA resource constraints play in the iterative refinements?}

While pinpointing results guide the accelerator refinement to move forward, resource constraints let programmers feed back to the design choice made in prior iterations.
Since it is difficult for programmers to foresee the resource consumption in future iterations, we assume that programmers will aggressively use resources in the current iteration, and look back to adjust resource distribution in future refinements.
To simplify the decision-making process, we make the following principle: If a strategy requires a moderate (less than 10\%) resource to implement, we will always satisfy this requirement.
The feedback then only occurs when multiple strategies all heavily consume the same type of resources.

The resources on FPGA fabric can be classified into two categories: computation resources---LUTs and DSPs, and memory resources (BRAM).
Table \ref{tab:resource} shows the resource consumption to realize each suggested strategies.
We can see that it is straightforward to make decisions on computation resource distribution, since only PE duplication heavily consumes LUTs and DSPs.
While almost all suggested strategies need BRAM, the design choice of the caching size is easy to make.
A 64KB BRAM buffer is sufficient to amortize the burst initial overhead, and costs only 2\% of the available BRAM resources (see Section \ref{subsec:cache_solution}).
Accordingly, implementing the double buffering strategy is also not a problem since it merely costs 3x BRAM blocks in caching three batches of data.

\begin{table}[h]
\centering
\caption{Summary of resource consumption of each strategy.}
\label{tab:resource}
{\scriptsize
\begin{tabular}{|l|l|l|}
\hline
Strategy & Compute Resource & Memory Resource \\ \hline \hline
Caching & Moderate & \begin{tabular}[c]{@{}l@{}}Heavy\\ Depends on Caching Size\end{tabular} \\ \hline
PE Duplication & \begin{tabular}[c]{@{}l@{}}Heavy\\ Depends on Dup Factor\end{tabular} & \begin{tabular}[c]{@{}l@{}}Heavy\\ Depends on Dup Factor\end{tabular} \\ \hline
Loop Pipelining & Moderate & Moderate \\ \hline
Double Buffering & Moderate & \begin{tabular}[c]{@{}l@{}}Heavy\\ $\sim$3x of Caching\end{tabular} \\ \hline
Scratchpad Reorg. & Moderate & \begin{tabular}[c]{@{}l@{}}Heavy\\ Depends on BRAM Width\end{tabular} \\ \hline
\end{tabular}
}
\end{table}

The feedback happens in the scratchpad reorganization iteration.
Apart from the BRAM blocks allocated for data caching, a computational kernel may need extra BRAM to store intermediate data, such as the dynamic programming kernels, NW and VITERBI, where a BRAM buffer is needed to store the bookkeeping information.
A large number of PE duplicates will consume a large portion of BRAM, leaving limited room for scratchpad reorganization.
As a result, there is a trade-off between the PE duplication factor and BRAM width.

This issue, however, does not affect the accelerator design process significantly.
First, only a few kernels, which need BRAM blocks apart from the ones allocated for explicit data caching, need to make this decision.
Moreover, the design space in this trade-off is very small, i.e., 16-bit, 32-bit, ..., up to 512-bit.
In fact, since a BRAM block can be reconfigured to 32-bit width, programmers only need to try at most four choices---64-bit, 128-bit, 256-bit and 512-bit, which is even affordably solved by a brute-force binary search.

In summary, in our proposed guideline, the iteration process is greatly simplified since no complex parameters tuning regarding the resource constraints is needed, and it achieves compelling results.
One may implement a more optimized accelerator that does not adhere to our guideline. 
However, the situation would become much more complicated, since there is an exponential design space of resource distribution to various resource-conflict optimization strategies.
Most prior work attempts to automatically search for the optimal solution for an individual strategy, e.g., automatic data tiling.
As an initial attempt, Wang et al.~\cite{fpga-opencl} also proposes a performance analysis framework to guide designers in choosing proper optimizations for OpenCL applications on Altera FPGAs. However, they still target hardware designers and do not present convincing speedups over CPU.
A comprehensive decision-making mechanism across all the strategies applied in an entire design flow is still an open problem; we leave this to future work.

\noindent \textbf{Q: Why does the paper compare the accelerator performance with that of a single CPU core?}

First, as mentioned earlier, FPGA is becoming a free bonus. That is, any speedup over a single CPU core is beneficial. Second, the speedups of our FPGA accelerators (except BFS and SPMV that is PCIe bound) are comparable or faster to the 12-core Xeon CPU we used even if the CPU got an ideal 12x speedup. Moreover, the FPGA only consumes roughly one tenth power of 12 cores, which is much more energy-efficient.



\noindent \textbf{Q: Given different speedups for different kernels, what category of kernels are better benefited by the CPU-FPGA platform and the proposed guideline?}

First, kernels that are computation-intensive have the potential to be accelerated by the platform.
The two kernels that do not achieve speedup are both communication-intensive kernels.
The MachSuite BFS kernel simply traverses all the nodes in a graph without doing any computation, and sparse matrix-vector multiplication is a well-known communication-intensive problem.
Such communication-intensive kernels often lead to a serious data transfer overhead compared to the CPU execution, which can probably be detected by the CPU-FPGA communication time measurement.

Moreover, kernels that conceive massive task-level parallelism can be accelerated.
The FPGA fabric consists of a great many distributed computation and memory building blocks and is naturally fit for applications with a large number of parallel tasks.
In addition to our study in this paper, prior work~\cite{blaze} has also integrated FPGA accelerators into the Hadoop and Spark MapReduce frameworks to accelerate MapReduce applications, e.g., DNA sequencing and data analytics.

Finally, our proposed guideline benefits kernels that spend a large portion of time on loop blocks.
Programmers can customize a hardware pipeline for a loop block through just a pragma, which allows multiple loop iterations to be executed simultaneously in a pipeline.
In addition, PE duplication that enables task-level parallelism can also be easily programmed through a pragma to unroll the loop with proper memory partitioning pragmas.

\noindent \textbf{Q: Does the proposed guideline free programmers from learning hardware knowledge in order to design FPGA accelerators?}

The proposed guideline does not let programmers completely get rid of hardware knowledge, but the learning curve is far shorter than that of evolving a software programmer into a conventional hardware accelerator design expert.
First, the proposed guideline applies the merit of data-driven iterative refinement methodology, which is commonly used by programmers in application performance optimization.
Second, the suggested strategies are all programmer-accessible, and cost merely $\sim$200 lines of HLS-C code to realize.
Programmers are required to collect particular architectural parameters in making design choices, which also commonly happens in application development.
In summary, the proposed guideline lowers the barrier to enter FPGA accelerator design and makes it accessible to mainstream software programmers.

\noindent \textbf{Q: What is the essential difference between a pure C program and a high-quality HLS-C accelerator design?}

The most significant difference is that explicit manipulation of the cache memory system is required in accelerator programming.
Both CPU and FPGA architectures require programmers to specify algorithms, but the former virtualize to programmers a low-latency memory system through hardware-supported cache hierarchy.
FPGAs, contrarily, supplies programmers with on-chip scratchpads that have more flexibility, but the fulfillment of this flexibility advantage requires explicit program statements.
As a consequence, almost all the suggested strategies are doing various on-chip scratchpad manipulations, e.g., explicit data caching, memory partitioning, double buffering and scratchpad reorganization.

\noindent \textbf{Q: Is it possible to automate the entire guideline into a push-button process through compilers?}

It is always the ultimate objective to make everything automated. While the paper delivers an encouraging message that a software programmer may also make high-quality FPGA accelerators without systematically learning RTL design expertise, the gap towards complete automation is still considerable. Most existing automation strategies, as we will illustrate in Section~\ref{sec:related}, focus on one optimization problem and more or less make some restrictions to user programs. It implies that these strategies are not only specific but hard to integrate together to form a comprehensive automation tool. For example, if we first apply an optimal auto-caching approach, which tries to utilize as many as memory to minimize computation latency, then other optimizations may have no room to perform. Although some developers attempt to build system-level automation frameworks~\cite{cmost,merlin}, the lack of a mature open source infrastructure and community prevents researchers from contributing their solutions to coordinate with others. Nonetheless, the proposed best practice guideline may serve as a direction for researchers and vendors to develop an effective automation flow that works for a broad class of applications.


\section{Related Work}
\label{sec:related}



\noindent \textbf{High-Level Synthesis Automation.} Many automated code transformations for the technologies covered in this paper have been proposed using commercial HLS tools or open-source tools such as LegUp~\cite{legup} and CHiMPS~\cite{chimps} as a back-end.

For \textit{on-chip data caching}, existing automation strategies mainly focus on analyzing data access patterns, identifying data reuse between loop instances, and then generating on-chip buffers with proper partitions~\cite{Cong-dac12, Pham-date15, Pouchet-fpga13}. However, most solutions only consider arrays with affine accesses. Automated data caching for an array with arbitrary (non-affine, random, or even both) access patterns is still an open research problem. For \textit{PE duplication}, the difficulty is that if we duplicate a large computation module, many hardware resources are required and imply less number of PEs. As a result, some researchers deal with this problem by developing algorithms to realize the duplication of a suitable PE granularity under resource constraints~\cite{Hagiescu-dac09, Cong-fpga14}, but leverage code modularization to users. Therefore, the restrictions on transforming user programs are still necessary. For \textit{pipelining}, the impediments to achieving fine-grained fully pipelining mainly include 1) data/loop-carried dependency, 2) uncertain loop bounds~\cite{liu-fccm15,liu-fccm16}, and 3) non-affine memory access~\cite{Venkat-pldi15}. Although many researchers have figured out some solutions to each problem, a complete solution is still missing. For \textit{double buffering}, the most widely used application is to form a coarse-grained (nested loop) pipeline~\cite{raghu-asplos,cgpa,elasticflow}. They extract necessary information from the problem using static analysis or user directives and apply predefined templates to form a coarse-grained pipeline using double buffers. Again, those solutions are not yet applicable to arbitrary user programs.

On the other hand, there have some advance techniques that highly rely on hardware expertise so we do not cover in this paper. For example, automated unified cache generation on FPGAs is implemented by \cite{raghu-asplos, weisz2013fpga}. Advanced on-chip memory partition optimizations to avoid data access conflicts for improving pipelining and parallelism are also well-studied~\cite{Wang-dac13,Su-fpga16,Cilardo-taco15}.


\noindent \textbf{Domain-Specific Languages.} While generating accelerators from generic programming languages presents challenges in discovering parallelism, pipeline structures and memory access, researchers have explored domain-specific languages~\cite{adrian2015} to describe certain patterns and structures using domain knowledge. Lime~\cite{lime} is a Java-based domain-specific language that provides several parallel patterns to improve the programmability. Bluespec~\cite{bluespec} is a functional hardware description language based on Haskell with atomic actions. Chisel~\cite{chisel} embeds hardware construction primitives with Scala and supports high-level abstractions and generators. DHDL~\cite{raghu-asplos,raghu-isca} is an intermediate hardware representation that can be generated from parallel patterns such as map, reduce, zip and filter. This dataflow representation can be used to generate low-level HDL and aid design space exploration. TABLA~\cite{tabla} provides an FPGA accelerator generator for machine learning algorithms that produces synthesizable Verilog code from user model specifications using a set of predesigned templates. Using a similar approach, DNNWeaver~\cite{dnnweaver} targets deep neural models. 

Although these tools can provide higher productivity and generate more efficient hardware when applications have certain amenable characteristics, they are often limited to small domains and do not work well for applications outside those domains. 



\section{Conclusion}
\label{sec:conclusion}

While the FPGA is changing its role from special-purpose hardware to primary computing resource, we demonstrate that mainstream software programmers can produce compelling FPGA accelerators with affordable programming efforts. 
For a wide class of applications from a state-of-the-art accelerator benchmark suite MachSuite, our best-effort guideline in HLS programming improves the naive accelerator performance by 42$\sim$29030x, which is up to 112.8x faster than a Xeon CPU core.
In our proposed best-effort guideline, we adopt a well-known data-driven iterative refinement methodology. 
During the refinement process, we apply five major programmer-friendly HLS optimization techniques, including explicit data caching, customized pipelining, processing element duplication, double buffering and bandwidth-aware scratchpad reorganization.
To provide more insights, for each optimization technique, we also quantitatively evaluate its performance impact and illustrate its software programming counterpart.
Although our best-effort guideline may not always produce the optimal accelerator, it is more accessible to software programmers and provides reasonable performance.
We hope this will stimulate more research in FPGA-based acceleration and facilitate its wide adoption in the software programming community. 


\bibliographystyle{ieeetr}
\bibliography{ref}

\end{document}